\documentclass[twocolumn]{aastex631}

\usepackage{soul}
\usepackage{color}
\usepackage{xspace}
\usepackage{graphicx}
\usepackage{amsmath}
\usepackage{amssymb}
\usepackage{yfonts}
\usepackage{xcolor}
\usepackage{float}
\hypersetup{pdfauthor={Name}}
\usepackage{stackengine}
\usepackage{xparse}
\usepackage{multirow}
\usepackage{longtable}
\usepackage{colortbl}
\usepackage{csquotes}
\usepackage{enumitem}
\usepackage{tikz}

\graphicspath{{./}{figures/}}

\shorttitle{GRB\,221009A}
\shortauthors{Lesage et al.}

\begin{document}

\title{Fermi-GBM Discovery of GRB 221009A:\\An Extraordinarily Bright GRB from Onset to Afterglow}

\correspondingauthor{S. Lesage}
\email{stephen.lesage@uah.edu}



\def\shrinkage{2.1mu}
\def\vecsign{\mathchar"017E}
\def\dvecsign{\smash{\stackon[-1.95pt]{\mkern-\shrinkage\vecsign}{\rotatebox{180}{$\mkern-\shrinkage\vecsign$}}}}
\def\dblvec#1{\def\useanchorwidth{T}\stackon[-4.2pt]{#1}{\,\dvecsign}}
\stackMath
\def\Avec{\vec{A}}
\def\Bvec{\vec{B}}
\def\Cvec{\vec{C}}
\def\Dvec{\vec{D}}
\def\Evec{\vec{E}}
\def\Fvec{\vec{F}}
\def\Gvec{\vec{G}}
\def\Hvec{\vec{H}}
\def\Ivec{\vec{I}}
\def\Jvec{\vec{J}}
\def\Kvec{\vec{K}}
\def\Lvec{\vec{L}}
\def\Mvec{\vec{M}}
\def\Nvec{\vec{N}}
\def\Ovec{\vec{O}}
\def\Pvec{\vec{P}}
\def\Qvec{\vec{Q}}
\def\Rvec{\vec{R}}
\def\Svec{\vec{S}}
\def\Tvec{\vec{T}}
\def\Uvec{\vec{U}}
\def\Vvec{\vec{V}}
\def\Wvec{\vec{W}}
\def\Xvec{\vec{X}}
\def\Yvec{\vec{Y}}
\def\Zvec{\vec{Z}}
\def\Ahat{\hat{A}}
\def\Bhat{\hat{B}}
\def\Chat{\hat{C}}
\def\Dhat{\hat{D}}
\def\Ehat{\hat{E}}
\def\Fhat{\hat{F}}
\def\Ghat{\hat{G}}
\def\Hhat{\hat{H}}
\def\Ihat{\hat{I}}
\def\Jhat{\hat{J}}
\def\Khat{\hat{K}}
\def\Lhat{\hat{L}}
\def\Mhat{\hat{M}}
\def\Nhat{\hat{N}}
\def\Ohat{\hat{O}}
\def\Phat{\hat{P}}
\def\Qhat{\hat{Q}}
\def\Rhat{\hat{R}}
\def\Shat{\hat{S}}
\def\That{\hat{T}}
\def\Uhat{\hat{U}}
\def\Vhat{\hat{V}}
\def\What{\hat{W}}
\def\Xhat{\hat{X}}
\def\Yhat{\hat{Y}}
\def\Zhat{\hat{Z}}
\def\Aten{\dblvec{A}}
\def\Bten{\dblvec{B}}
\def\Cten{\dblvec{C}}
\def\Dten{\dblvec{D}}
\def\Eten{\dblvec{E}}
\def\Ften{\dblvec{F}}
\def\Gten{\dblvec{G}}
\def\Hten{\dblvec{H}}
\def\Iten{\dblvec{I}}
\def\Jten{\dblvec{J}}
\def\Kten{\dblvec{K}}
\def\Lten{\dblvec{L}}
\def\Mten{\dblvec{M}}
\def\Nten{\dblvec{N}}
\def\Oten{\dblvec{O}}
\def\Pten{\dblvec{P}}
\def\Qten{\dblvec{Q}}
\def\Rten{\dblvec{R}}
\def\Sten{\dblvec{S}}
\def\Tten{\dblvec{T}}
\def\Uten{\dblvec{U}}
\def\Vten{\dblvec{V}}
\def\Wten{\dblvec{W}}
\def\Xten{\dblvec{X}}
\def\Yten{\dblvec{Y}}
\def\Zten{\dblvec{Z}}
\def\Aol{$\mathbbm{A}$}
\def\Bol{$\mathbbm{B}$}
\def\Col{$\mathbbm{C}$}
\def\Dol{$\mathbbm{D}$}
\def\Eol{$\mathbbm{E}$}
\def\Fol{$\mathbbm{F}$}
\def\Gol{$\mathbbm{G}$}
\def\Hol{$\mathbbm{H}$}
\def\Iol{$\mathbbm{I}$}
\def\Jol{$\mathbbm{J}$}
\def\Kol{$\mathbbm{K}$}
\def\Lol{$\mathbbm{L}$}
\def\Mol{$\mathbbm{M}$}
\def\Nol{$\mathbbm{N}$}
\def\Ool{$\mathbbm{O}$}
\def\Pol{$\mathbbm{P}$}
\def\Qol{$\mathbbm{Q}$}
\def\Rol{$\mathbbm{R}$}
\def\Sol{$\mathbbm{S}$}
\def\Tol{$\mathbbm{T}$}
\def\Uol{$\mathbbm{U}$}
\def\Vol{$\mathbbm{V}$}
\def\Wol{$\mathbbm{W}$}
\def\Xol{$\mathbbm{X}$}
\def\Yol{$\mathbbm{Y}$}
\def\Zol{$\mathbbm{Z}$}
\def\avec{\vec{a}}
\def\bvec{\vec{b}}
\def\cvec{\vec{c}}
\def\dvec{\vec{d}}
\def\evec{\vec{e}}
\def\fvec{\vec{f}}
\def\gvec{\vec{g}}
\def\hvec{\vec{h}}
\def\ivec{\vec{i}}
\def\jvec{\vec{j}}
\def\kvec{\vec{k}}
\def\lvec{\vec{l}}
\def\mvec{\vec{m}}
\def\nvec{\vec{n}}
\def\ovec{\vec{o}}
\def\pvec{\vec{p}}
\def\qvec{\vec{q}}
\def\rvec{\vec{r}}
\def\svec{\vec{s}}
\def\tvec{\vec{t}}
\def\uvec{\vec{u}}
\def\vvec{\vec{v}}
\def\wvec{\vec{w}}
\def\xvec{\vec{x}}
\def\yvec{\vec{y}}
\def\zvec{\vec{z}}
\def\ahat{\hat{a}}
\def\bhat{\hat{b}}
\def\chat{\hat{c}}
\def\dhat{\hat{d}}
\def\ehat{\hat{e}}
\def\fhat{\hat{f}}
\def\ghat{\hat{g}}
\def\hhat{\hat{h}}
\def\ihat{\hat{i}}
\def\jhat{\hat{j}}
\def\khat{\hat{k}}
\def\lhat{\hat{l}}
\def\mhat{\hat{m}}
\def\nhat{\hat{n}}
\def\ohat{\hat{o}}
\def\phat{\hat{p}}
\def\qhat{\hat{q}}
\def\rhat{\hat{r}}
\def\shat{\hat{s}}
\def\that{\hat{t}}
\def\uhat{\hat{u}}
\def\vhat{\hat{v}}
\def\what{\hat{w}}
\def\xhat{\hat{x}}
\def\yhat{\hat{y}}
\def\zhat{\hat{z}}
\def\aten{\dblvec{a}}
\def\bten{\dblvec{b}}
\def\cten{\dblvec{c}}
\def\dten{\dblvec{d}}
\def\eten{\dblvec{e}}
\def\ften{\dblvec{f}}
\def\gten{\dblvec{g}}
\def\hten{\dblvec{h}}
\def\iten{\dblvec{i}}
\def\jten{\dblvec{j}}
\def\kten{\dblvec{k}}
\def\lten{\dblvec{l}}
\def\mten{\dblvec{m}}
\def\nten{\dblvec{n}}
\def\oten{\dblvec{o}}
\def\pten{\dblvec{p}}
\def\qten{\dblvec{q}}
\def\rten{\dblvec{r}}
\def\sten{\dblvec{s}}
\def\tten{\dblvec{t}}
\def\uten{\dblvec{u}}
\def\vten{\dblvec{v}}
\def\wten{\dblvec{w}}
\def\xten{\dblvec{x}}
\def\yten{\dblvec{y}}
\def\zten{\dblvec{z}}
\def\aol{$\mathbbm{a}$}
\def\bol{$\mathbbm{b}$}
\def\col{$\mathbbm{c}$}
\def\dol{$\mathbbm{d}$}
\def\eol{$\mathbbm{e}$}
\def\fol{$\mathbbm{f}$}
\def\gol{$\mathbbm{g}$}
\def\hol{$\mathbbm{h}$}
\def\iol{$\mathbbm{i}$}
\def\jol{$\mathbbm{j}$}
\def\kol{$\mathbbm{k}$}
\def\lol{$\mathbbm{l}$}
\def\mol{$\mathbbm{m}$}
\def\nol{$\mathbbm{n}$}
\def\ool{$\mathbbm{o}$}
\def\pol{$\mathbbm{p}$}
\def\qol{$\mathbbm{q}$}
\def\rol{$\mathbbm{r}$}
\def\sol{$\mathbbm{s}$}
\def\tol{$\mathbbm{t}$}
\def\uol{$\mathbbm{u}$}
\def\vol{$\mathbbm{v}$}
\def\wol{$\mathbbm{w}$}
\def\xol{$\mathbbm{x}$}
\def\yol{$\mathbbm{y}$}
\def\zol{$\mathbbm{z}$}
\newcommand{\eps}{\epsilon}
\newcommand{\veps}{\varepsilon}
\newcommand{\vtheta}{\vartheta}
\newcommand{\vphi}{\varphi}
\newcommand{\vrho}{\varrho}
\def\alphavec{\vec{\alpha}}
\def\nuvec{\vec{\nu}}
\def\betavec{\vec{\beta}}
\def\xivec{\vec{\xi}}
\def\Xivec{\vec{\Xi}}
\def\gammavec{\vec{\gamma}} 
\def\Gammavec{\vec{\Gamma}}
\def\deltavec{\vec{\delta}} 
\def\Deltavec{\vec{\Delta}}
\def\pivec{\vec{\pi}} 
\def\Pivec{\vec{\Pi}}
\def\epsvec{\vec{\eps}} 
\def\vepsvec{\vec{\veps}} 
\def\rhovec{\vec{\rho}}
\def\vrhovec{\vec{\vrho}}
\def\zetavec{\vec{\zeta}}
\def\sigmavec{\vec{\sigma}}
\def\Sigmavec{\vec{\Sigma}}
\def\etavec{\vec{\eta}}
\def\tauvec{\vec{\tau}}
\def\thetavec{\vec{\theta}}
\def\vthetavec{\vec{\vtheta}}
\def\Thetavec{\vec{\Theta}}
\def\upsilonvec{\vec{\upsilon}}
\def\Upsilonvec{\vec{\Upsilon}}
\def\iotavec{\vec{\iota}}
\def\phivec{\vec{\phi}}
\def\vphivec{\vec{\vphi}}
\def\Phivec{\vec{\Phi}}
\def\kappavec{\vec{\kappa}}
\def\chivec{\vec{\chi}}
\def\lambdavec{\vec{\lambda}}
\def\Lambdavec{\vec{\Lambda}}
\def\psivec{\vec{\psi}}
\def\Psivec{\vec{\Psi}}
\def\muvec{\vec{\mu}}
\def\omegavec{\vec{\omega}}
\def\Omegavec{\vec{\Omega}}
\def\alphahat{\hat{\alpha}}
\def\nuhat{\hat{\nu}}
\def\betahat{\hat{\beta}}
\def\xihat{\hat{\xi}}
\def\Xihat{\hat{\Xi}}
\def\gammahat{\hat{\gamma}} 
\def\Gammahat{\hat{\Gamma}}
\def\deltahat{\hat{\delta}} 
\def\Deltahat{\hat{\Delta}}
\def\pihat{\hat{\pi}} 
\def\Pihat{\hat{\Pi}}
\def\epshat{\hat{\eps}} 
\def\vepshat{\hat{\veps}} 
\def\rhohat{\hat{\rho}}
\def\vrhohat{\hat{\vrho}}
\def\zetahat{\hat{\zeta}}
\def\sigmahat{\hat{\sigma}}
\def\Sigmahat{\hat{\Sigma}}
\def\etahat{\hat{\eta}}
\def\tauhat{\hat{\tau}}
\def\thetahat{\hat{\theta}}
\def\vthetahat{\hat{\vtheta}}
\def\Thetahat{\hat{\Theta}}
\def\upsilonhat{\hat{\upsilon}}
\def\Upsilonhat{\hat{\Upsilon}}
\def\iotahat{\hat{\iota}}
\def\phihat{\hat{\phi}}
\def\vphihat{\hat{\vphi}}
\def\Phihat{\hat{\Phi}}
\def\kappahat{\hat{\kappa}}
\def\chihat{\hat{\chi}}
\def\lambdahat{\hat{\lambda}}
\def\Lambdahat{\hat{\Lambda}}
\def\psihat{\hat{\psi}}
\def\Psihat{\hat{\Psi}}
\def\muhat{\hat{\mu}}
\def\omegahat{\hat{\omega}}
\def\Omegahat{\hat{\Omega}}
\def\alphaten{\dblvec{\alpha}}
\def\nuten{\dblvec{\nu}}
\def\betaten{\dblvec{\beta}}
\def\xiten{\dblvec{\xi}}
\def\Xiten{\dblvec{\Xi}}
\def\gammaten{\dblvec{\gamma}} 
\def\Gammaten{\dblvec{\Gamma}}
\def\deltaten{\dblvec{\delta}} 
\def\Deltaten{\dblvec{\Delta}}
\def\piten{\dblvec{\pi}} 
\def\Piten{\dblvec{\Pi}}
\def\epsten{\dblvec{\eps}} 
\def\vepsten{\dblvec{\veps}} 
\def\rhoten{\dblvec{\rho}}
\def\vrhoten{\dblvec{\vrho}}
\def\zetaten{\dblvec{\zeta}}
\def\sigmaten{\dblvec{\sigma}}
\def\Sigmaten{\dblvec{\Sigma}}
\def\etaten{\dblvec{\eta}}
\def\tauten{\dblvec{\tau}}
\def\thetaten{\dblvec{\theta}}
\def\vthetaten{\dblvec{\vtheta}}
\def\Thetaten{\dblvec{\Theta}}
\def\upsilonten{\dblvec{\upsilon}}
\def\Upsilonten{\dblvec{\Upsilon}}
\def\iotaten{\dblvec{\iota}}
\def\phiten{\dblvec{\phi}}
\def\vphiten{\dblvec{\vphi}}
\def\Phiten{\dblvec{\Phi}}
\def\kappaten{\dblvec{\kappa}}
\def\chiten{\dblvec{\chi}}
\def\lambdaten{\dblvec{\lambda}}
\def\Lambdaten{\dblvec{\Lambda}}
\def\psiten{\dblvec{\psi}}
\def\Psiten{\dblvec{\Psi}}
\def\muten{\dblvec{\mu}}
\def\omegaten{\dblvec{\omega}}
\def\Omegaten{\dblvec{\Omega}}
\def\alphaol{$\mathbb{\alpha}$}
\def\nuol{$\mathbb{\nu}$}
\def\betaol{$\mathbb{\beta}$}
\def\xiol{$\mathbb{\xi}$}
\def\Xiol{$\mathbb{\Xi}$}
\def\gammaol{$\mathbb{\gamma}$}
\def\Gammaol{$\mathbb{\Gamma}$}
\def\deltaol{$\mathbb{\delta}$}
\def\Deltaol{$\mathbb{\Delta}$}
\def\piol{$\mathbb{\pi}$}
\def\Piol{$\mathbb{\Pi}$}
\def\epsol{$\mathbb{\eps}$}
\def\vepsol{$\mathbb{\veps}$}
\def\rhool{$\mathbb{\rho}$}
\def\vrhool{$\mathbb{\vrho}$}
\def\zetaol{$\mathbb{\zeta}$}
\def\sigmaol{$\mathbb{\sigma}$}
\def\Sigmaol{$\mathbb{\Sigma}$}
\def\etaol{$\mathbb{\eta}$}
\def\tauol{$\mathbb{\tau}$}
\def\thetaol{$\mathbb{\theta}$}
\def\vthetaol{$\mathbb{\vtheta}$}
\def\Thetaol{$\mathbb{\Theta}$}
\def\upsilonol{$\mathbb{\upsilon}$}
\def\Upsilonol{$\mathbb{\Upsilon}$}
\def\iotaol{$\mathbb{\iota}$}
\def\phiol{$\mathbb{\phi}$}
\def\vphiol{$\mathbb{\vphi}$}
\def\Phiol{$\mathbb{\Phi}$}
\def\kappaol{$\mathbb{\kappa}$}
\def\chiol{$\mathbb{\chi}$}
\def\lambdaol{$\mathbb{\lambda}$}
\def\Lambdaol{$\mathbb{\Lambda}$}
\def\psiol{$\mathbb{\psi}$}
\def\Psiol{$\mathbb{\Psi}$}
\def\muol{$\mathbb{\mu}$}
\def\omegaol{$\mathbb{\omega}$}
\def\Omegaol{$\mathbb{\Omega}$}
\def\cross{\times}
\def\dot{\cdot}
\def\del{\nabla}
\def\delcross{\nabla \times}
\def\deldot{\nabla \cdot}
\def\delsq{\nabla^2}
\newcommand{\rarrow}{\Rightarrow}
\newcommand{\rrarrow}{\Longrightarrow}
\newcommand{\larrow}{\Leftarrow}
\newcommand{\llarrow}{\Longleftarrow}
\newcommand{\lrarrow}{\Leftrightarrow}
\newcommand{\llrrarrow}{\iff}
\newcommand{\nperp}{\not\perp}
\let\oldinf\inf
\renewcommand{\inf}{\infty}
\def\wbox{\square}
\def\bbox{\blacksquare}
\def\deg{^{\circ}}
\newcommand{\trm}{\textrm}
\newcommand{\tbf}{\textbf}
\newcommand{\tul}{\underline}
\newcommand{\tit}{\textit}
\newcommand{\texp}[1]{$^{\textrm{#1}}$}
\newcommand{\tqu}{\enquote}
\newcommand{\pref}[1]{(\pageref{#1})}
\newcommand{\eref}[1]{equation \eqref{#1}}
\newcommand{\avg}[1]{\overline{#1}}
\newcommand{\p}[1]{\left( #1 \right)}
\newcommand{\pp}[1]{\left[ #1 \right]}
\newcommand{\psqu}[1]{\left\{ #1 \right\}}
\newcommand{\pang}[1]{\left\langle #1 \right\rangle}
\newcommand{\abs}[1]{\left| #1 \right|}
\newcommand{\dabs}[1]{\left\lVert #1 \right\rVert}
\newcommand{\eval}[2]{\rvert_{#1}^{#2}}
\newcommand{\Eval}[2]{\Bigg\rvert_{#1}^{#2}}
\newcommand{\e}[1]{\times 10^{#1}}
\newcommand{\dv}[2]{\frac{d #1}{d #2}}
\newcommand{\ndv}[3]{\frac{d^{#1} #2}{d #3^{#1}}}
\newcommand{\pdv}[2]{\frac{\partial #1}{\partial #2}}
\newcommand{\npdv}[3]{\frac{\partial^{#1} #2}{\partial #3^{#1}}}
\newcommand{\ulabel}[2]{\underset{\mathclap{\substack{\uparrow\\#2}}}{#1}}
\newcommand{\llabel}[2]{\overset{\mathclap{\substack{#2\\\downarrow}}}{#1}}
\newcommand{\ublabel}[2]{\overbrace{#1}^{\mathclap{\substack{#2}}}}
\newcommand{\lblabel}[2]{\underbrace{#1}_{\mathclap{\substack{#2}}}}
\newcommand{\uslabel}[2]{\overbracket{#1}^{\mathclap{\substack{#2}}}}
\newcommand{\lslabel}[2]{\underbracket{#1}_{\mathclap{\substack{#2}}}}
\let\oldlim\lim
\renewcommand{\lim}[2]{\oldlim\limits_{{#1} \rightarrow {#2}}}
\let\oldsum\sum
\renewcommand{\sum}[2]{\oldsum\limits_{#1}^{#2}}
\let\oldprod\prod
\renewcommand{\prod}[2]{\oldprod\limits_{#1}^{#2}}
\let\oldint\int
\renewcommand{\int}[2]{\oldint\limits_{#1}^{#2}}
\newcommand{\dint}[4]{\oldint\limits_{#1}^{#2} \oldint\limits_{#3}^{#4}}
\newcommand{\tint}[6]{\oldint\limits_{#1}^{#2} \oldint\limits_{#3}^{#4} \oldint\limits_{#5}^{#6}}
\def\lint{\int_{l}}
\def\sint{\iint\limits_{S}}
\def\vint{\iiint\limits_{V}}
\def\olint{\oint\limits_{l}}
\def\osint{\oiint\limits_{S}}
\def\ovint{\oiint\limits_{V}\hspace{-10.9pt} \oldint_{}^{}}
\newcommand{\eq}[1]{\begin{equation*} #1 \end{equation*}}
\newcommand{\eql}[2]{\begin{equation} \label{#1} #2 \end{equation}}
\newcommand{\teq}[1]{$ #1 $}
\let\oldmatrix\matrix
\renewcommand{\matrix}[1]{$\begin{pmatrix} #1 \end{pmatrix}$}


\def\nm{\mbox{~nm}} 
\def\mum{\mbox{~\mu\hbox{m}}} 
\def\mm{\mbox{~mm}} 
\def\cm{\mbox{~cm}} 
\def\m{\mbox{~m}} 
\def\km{\mbox{~km}} 
\def\pc{\mbox{~pc}} 
\def\kpc{\mbox{~kpc}} 
\def\Mpc{\mbox{~Mpc}} 
\def\Gpc{\mbox{~Gpc}} 
\def\erg{\mbox{~erg}}
\def\eV{\mbox{~eV}} 
\def\keV{\mbox{~keV}} 
\def\MeV{\mbox{~MeV}} 
\def\GeV{\mbox{~GeV}} 
\def\TeV{\mbox{~TeV}} 
\def\Hz{\mbox{~Hz}} 
\def\kHz{\mbox{~kHz}} 
\def\MHz{\mbox{~MHz}} 
\def\GHz{\mbox{~GHz}} 
\def\THz{\mbox{~THz}} 
\def\ns{\mbox{~ns}} 
\def\mus{~\mu\hbox{s}} 
\def\ms{\mbox{~ms}} 
\def\sec{\mbox{~s}} 
\def\min{\mbox{~m}}
\def\hr{\mbox{~h}}
\def\yr{\mbox{~yr}}
\def\Jy{\mbox{~Jy}}
\def\ms{{\mbox{~ms}}}
\def\astar{A$_{\star}$}
\def\Msun{M$_{\odot}$}
\def\swift{{\textit{Swift}}\xspace}
\def\fermi{{\textit{Fermi}}\xspace}
\def\gbm{{\textit{Fermi}-GBM}\xspace}
\def\lat{{\textit{Fermi}-LAT}\xspace}
\def\Epk{E$_{\textrm{peak}}$}
\def\Eiso{E$_{\textrm{iso}}$}
\def\Liso{L$_{\textrm{iso}}$}
\def\t90{T$_{\textrm{90}}$}
\def\tvar{$t_{\textrm{var}}$}
\def\t0{$t_{0}$}
\def\nufnu{$\nu F_{\nu}$}
\def\ra#1#2#3{#1$^{^\textrm{h}}$#2$^{^\textrm{m}}$#3$^{^\textrm{s}}$}
\def\dec#1#2#3{#1$^\circ$#2$'$#3$''$}
\def\fluence{\textrm{erg}\cdot\textrm{cm}^{-2}}


\definecolor{burntorange}{rgb}{0.8, 0.33, 0.0}
\definecolor{amber}{rgb}{1.0, 0.75, 0.0}
\definecolor{ao(english)}{rgb}{0.0, 0.5, 0.0}
\definecolor{darkorchid}{rgb}{0.6, 0.2, 0.8}
\definecolor{aqua}{rgb}{0.0, 1.0, 1.0}
\definecolor{brightlavender}{rgb}{0.75, 0.58, 0.89}
\definecolor{lightgray}{gray}{0.9}
\definecolor{darkgray}{gray}{0.7}

\newcommand{\red}[1]{\textcolor{red}{#1}}
\newcommand{\orange}[1]{\textcolor{burntorange}{#1}}
\newcommand{\yellow}[1]{\textcolor{amber}{#1}}
\newcommand{\green}[1]{\textcolor{ao(english)}{#1}}
\newcommand{\blue}[1]{\textcolor{blue}{#1}}
\newcommand{\purple}[1]{\textcolor{darkorchid}{#1}}

\newcommand{\hlr}[1]{\sethlcolor{pink}{\hl{#1}}\sethlcolor{yellow}}
\newcommand{\hlo}[1]{\sethlcolor{orange}{\hl{#1}}\sethlcolor{yellow}}
\newcommand{\hly}[1]{\sethlcolor{yellow}{\hl{#1}}\sethlcolor{yellow}}
\newcommand{\hlg}[1]{\sethlcolor{green}{\hl{#1}}\sethlcolor{yellow}}
\newcommand{\hlb}[1]{\sethlcolor{aqua}{\hl{#1}}\sethlcolor{yellow}}
\newcommand{\hlp}[1]{\sethlcolor{brightlavender}{\hl{#1}}\sethlcolor{yellow}}

\NewDocumentCommand\bullets{>{\SplitList{;}}m}
  {
    \begin{itemize}
      \ProcessList{#1}{ \insertitem }
    \end{itemize}
  }
\NewDocumentCommand\numbers{>{\SplitList{;}}m}
  {
    \begin{enumerate}
      \ProcessList{#1}{ \insertitem }
    \end{enumerate}
  }
\newcommand\insertitem[1]{\item #1}

\def\pom{$\pm$\xspace}
\def\abt{$\sim$}

\newcommand{\kTmin}{$kT_{\textrm{min}}$\xspace}
\newcommand{\kTmax}{$kT_{\textrm{max}}$\xspace}
\newcommand{\Cstat}{$C_{\textrm{stat}}$\xspace}
\newcommand{\PGstat}{$PG_{\textrm{stat}}$\xspace}
\newcommand{\Pstat}{$P_{\textrm{stat}}$\xspace}
\newcommand{\chisq}{$\chi^{2}$\xspace}
\newcommand{\chisqdof}{$\chi^{2}_{\nu}$\xspace}

\def\checkmark{\tikz\fill[scale=0.4](0,.35) -- (.25,0) -- (1,.7) -- (.25,.15) -- cycle;} 

\newcommand{\tte}{\texttt{TTE}\xspace}
\newcommand{\ctime}{\texttt{CTIME}\xspace}
\newcommand{\cspec}{\texttt{CSPEC}\xspace}
\newcommand{\drm}{\texttt{DRM}\xspace}
\newcommand{\drms}{\texttt{DRM}s\xspace}
\newcommand{\lle}{\texttt{LLE}\xspace}
\def\targeted{\texttt{targeted search}\xspace}
\def\untargeted{\texttt{untargeted search}\xspace}


\newcommand{\grb}{GRB 221009A\xspace}

\setlength{\LTcapwidth}{\textwidth}


\def\aa{A\&A} 
\def\aar{A\&A~Rev.} 
\def\aas{A\&AS} 
\def\actaa{Acta Astron.} 
\def\aj{AJ} 
\def\ao{Appl.~Opt.} 
\def\apj{ApJ} 
\def\apjl{ApJ} 
\def\apjs{ApJS} 
\def\apjsupp{ApJ Supp. Series} 
\def\aplett{Astrophys.~Lett.} 
\def\apspr{Astrophys.~Space~Phys.~Res.} 
\def\apss{Ap\&SS} 
\def\araa{ARA\&A} 
\def\arxiv{arXiv} 
\def\arxive{arXiv e-prints} 
\def\azh{AZh} 
\def\baas{BAAS} 
\def\bac{Bull. astr. Inst. Czechosl.} 
\def\bain{Bull.~Astron.~Inst.~Netherlands} 
\def\caa{Chinese Astron. Astrophys.} 
\def\cjaa{Chinese J. Astron. Astrophys.} 
\def\cqg{CQGrav} 
\def\ea{Exp. Astron.} 
\def\fcp{Fund.~Cosmic~Phys.} 
\def\gca{Geochim.~Cosmochim.~Acta} 
\def\gcn{GCN Circ. } 
\def\grl{Geophys.~Res.~Lett.} 
\def\iaucirc{IAU~Circ.} 
\def\icarus{Icarus} 
\def\jcap{J. Cosmology Astropart. Phys.} 
\def\jcp{J.~Chem.~Phys.} 
\def\jgr{J.~Geophys.~Res.} 
\def\jqsrt{J.~Quant.~Spec.~Radiat.~Transf.} 
\def\jrasc{JRASC} 
\def\memras{MmRAS} 
\def\memsai{Mem.~Soc.~Astron.~Italiana} 
\def\mnras{MNRAS} 
\def\na{New A} 
\def\nar{New A Rev.} 
\def\nat{Nature} 
\def\nphysa{Nucl.~Phys.~A} 
\def\nucinstrummethodsa{Nucl.~Instrum.~Methods~A} 
\def\pasa{PASA} 
\def\pasj{PASJ} 
\def\pasp{PASP} 
\def\physrep{Phys.~Rep.} 
\def\physscr{Phys.~Scr} 
\def\planss{Planet.~Space~Sci.} 
\def\pra{Phys.~Rev.~A} 
\def\prb{Phys.~Rev.~B} 
\def\prc{Phys.~Rev.~C} 
\def\prd{Phys.~Rev.~D} 
\def\pre{Phys.~Rev.~E} 
\def\prx{Phys.~Rev.~X} 
\def\prl{Phys.~Rev.~Lett.} 
\def\procspie{Proc.~SPIE} 
\def\psci{PoS} 
\def\qjras{QJRAS} 
\def\rmxaa{Rev. Mexicana Astron. Astrofis.} 
\def\skytel{S\&T} 
\def\solphys{Sol.~Phys.} 
\def\ssr{Space~Sci.~Rev.} 
\def\sovast{Soviet~Ast.} 
\def\zap{ZAp} 


\newcommand\aastex{AAS\TeX}
\newcommand\latex{La\TeX}

%
\newcommand{\CNRS}{\affiliation{Universit\'e Paris-Saclay, CNRS/IN2P3, IJCLab, 91405 Orsay, France}}
\newcommand{\CSPAR}{\affiliation{Center for Space Plasma and Aeronomic Research, University of Alabama in Huntsville, Huntsville, AL 35899, USA}}
\newcommand{\FIT}{\affiliation{Department of Aerospace, Physics and Space Sciences, Florida Institute of Technology, Melbourne, FL 32901, USA}}
\newcommand{\GSFC}{\affiliation{NASA Goddard Space Flight Center, University of Maryland, Baltimore County, Greenbelt, MD 20771, USA}}
\newcommand{\GSFCAstro}{\affiliation{Astrophysics Science Division, NASA Goddard Space Flight Center, Greenbelt, MD 20771, USA}}
\newcommand{\ISSI}{\affiliation{International Space Science Institute, Hallerstrasse 6, 3012 Bern, Switzerland}}
\newcommand{\Jacobs}{\affiliation{Jacobs Space Exploration Group, Huntsville, AL 35806, USA}}
\newcommand{\LSU}{\affiliation{Department of Physics and Astronomy, Louisiana State University, Baton Rouge, LA 70803 USA}}
\newcommand{\LosAlamos}{\affiliation{Center for Non Linear Studies, Los Alamos National Laboratory, Los Alamos, NM, 87545, USA}}
\newcommand{\MPI}{\affiliation{Max-Planck-Institut f\"{u}r extraterrestrische Physik, Giessenbachstrasse 1, D-85748 Garching, Germany}}
\newcommand{\MSFCAstro}{\affiliation{ST12 Astrophysics Branch, NASA Marshall Space Flight Center, Huntsville, AL 35812, USA}}
\newcommand{\NASA}{\altaffiliation{NASA Postdoctoral Fellow}}
\newcommand{\NYUAbuDhabi}{\affiliation{Center for Astro, Particle, and Planetary Physics, New York University Abu Dhabi}}
\newcommand{\PdiB}{\affiliation{Dipartimento Interateneo di Fisica dell'Università e Politecnico di Bari, Via E. Orabona 4, 70125, Bari, Italy}}
\newcommand{\SdiB}{\affiliation{Istituto Nazionale di Fisica Nucleare - Sezione di Bari, Via E. Orabona 4, 70125, Bari, Italy}}
\newcommand{\SPA}{\affiliation{Department of Space Science, University of Alabama in Huntsville, 320 Sparkman Drive, Huntsville, AL 35899, USA}}
\newcommand{\SRON}{\affiliation{SRON Netherlands Institute for Space Research, Niels Bohrweg 4, 2333CA Leiden, The Netherlands}}
\newcommand{\UAH}{\affiliation{University of Alabama in Huntsville, 320 Sparkman Drive, Huntsville, AL 35899, USA}}
\newcommand{\UCD}{\affiliation{School of Physics, University College Dublin, Belfield, Dublin 4, Ireland}}
\newcommand{\USRA}{\affiliation{Science and Technology Institute, Universities Space Research Association, Huntsville, AL 35805, USA}}
%

%
\author[0000-0001-8058-9684]{S.~Lesage}
\SPA
\CSPAR
\author[0000-0002-2149-9846]{P.~Veres}
\SPA
\CSPAR
\author[0000-0003-2105-7711]{M.~S.~Briggs}
\SPA
\CSPAR
\author[0000-0002-0587-7042]{A.~Goldstein}
\USRA
\author[0000-0001-9201-4706]{D.~Kocevski}
\MSFCAstro
\author[0000-0002-2942-3379]{E.~Burns}
\LSU
\author[0000-0002-8585-0084]{C.~A.~Wilson-Hodge}
\MSFCAstro
\author[0000-0001-7916-2923]{P.~N.~Bhat}
\SPA
\CSPAR
\author[0000-0002-1169-7486]{D.~Huppenkothen}
\SRON
\author[0000-0003-2624-0056]{C.~L.~Fryer}
\LosAlamos
% ...



\begin{abstract}

We report the discovery of \grb, the highest flux gamma-ray burst ever observed by the \fermi Gamma-ray Burst Monitor (\gbm). This GRB has continuous prompt emission lasting more than 600 seconds which smoothly transitions to afterglow visible in the \gbm energy range (8\,keV--40\,MeV), and total energetics higher than any other burst in the \gbm sample. By using a variety of new and existing analysis techniques we probe the spectral and temporal evolution of \grb. We find no emission prior to the \gbm trigger time (\t0; 2022 October 9 at 13:16:59.99 UTC), indicating that this is the time of prompt emission onset. The triggering pulse exhibits distinct spectral and temporal properties suggestive of the thermal, photospheric emission of shock-breakout, with significant emission up to $\sim$15\,MeV. We characterize the onset of external shock at \t0+600\,s and find evidence of a plateau region in the early-afterglow phase which transitions to a slope consistent with \swift-XRT afterglow measurements. We place the total energetics of \grb in context with the rest of the \gbm sample and find that this GRB has the highest total isotropic-equivalent energy ($\textrm{E}_{\gamma,\textrm{iso}}=1.0\times10^{55}$\,erg) and second highest isotropic-equivalent luminosity ($\textrm{L}_{\gamma,\textrm{iso}}=9.9\times10^{53}$\,erg/s) based on redshift of z = 0.151. These extreme energetics are what allowed us to observe the continuously emitting central engine of \gbm from the beginning of the prompt emission phase through the onset of early afterglow.

\end{abstract}
\keywords{gamma rays: individual (221009A)}

\section{Introduction} \label{sec:introduction}

Gamma-ray bursts (GRBs) are the brightest signatures of stellar deaths in the Universe. They are produced by ultra-relativistic, collimated jets formed via two main progenitor channels: the merging of binary neutron star systems (BNSs; \citealt{Eichler1989, Narayan1992, Abbott2017a}) and the core collapse of rapidly-rotating massive, stripped-envelope stars \citep{Woosley1993, MacFadyen1999}. Prompt GRB emission is highly variable, lasting from tens of milliseconds up to ten thousand seconds, and is thought to arise from either internal shocks or magnetic reconnections within a relativistically expanding jet \citep{Goodman1986, Paczynski1986, Rees1994}. The prompt emission is followed by a longer-lived external shock \enquote{afterglow} component that arises from the interaction of the jet with the surrounding medium.  This afterglow component can be observed across the electromagnetic spectrum and is well-described as non-thermal synchrotron radiation arising from shock-accelerated electrons leading the jet \citep{Meszaros1993, Meszaros1997, Sari1998}. 

The Gamma-ray Burst Monitor onboard the \textit{Fermi Gamma-ray Space Telescope} (\gbm) has detected over 3500 GRBs, with approximately 8\% of these bursts also detected by the \fermi Large Area Telescope (\lat) at energies $>$100\,MeV \citep{Ajello2019}. The resulting observations have dramatically expanded our current understanding of broadband emission from GRBs, yet fundamental questions regarding the physical processes and radiation mechanisms that produce the GRB prompt emission still remain.

Emitting the bulk of their energy between 1\,keV and 1\,MeV, GRBs exhibit a range of spectral shapes and can include multiple components, posing significant challenges to attributing the prompt emission to any single physical process or radiation mechanism. The broadband prompt emission is characterized as a featureless non-thermal spectrum that is most commonly modeled with a phenomenological smoothly broken power-law model known as the Band function \citep{Band1993}. This emission is typically attributed to optically thin synchrotron emission, despite long-standing challenges in matching model predictions to the observed spectral shape of the prompt emission \citep{Preece1998}. Detailed time-resolved analysis of \gbm observations have also revealed additional low-energy components superimposed on the broadband spectra, including a quasi-thermal component thought to be due to emission from an optically thick photosphere \citep{Ryde2011, Guiriec2011, Axelsson2012, Guiriec2015}. Studies have shown that some of the difficulties in attributing the shape of the GRB prompt emission to synchrotron emission can be alleviated by the inclusion of these additional low-energy components \citep{Burgess2011, Guiriec2011, Beniamini2013, Oganesyan2017}. 

Deciphering GRB spectra is further complicated by the presence (or lack thereof) of high-energy emission ($\gtrsim$100\,MeV). For GRBs observed by both the \gbm and \lat, an additional high-energy power-law is often required to fit emission extending above $\sim$100\,MeV, and evidence for such a component can often be seen extending down into X-ray energies ($< 20$\,keV). The emergence of this component is typically delayed with respect to the initial keV emission and has been attributed to afterglow emission arising during the prompt phase \citep{Abdo2009, Ackermann2010, Ackermann2011, Ackermann2014}. A higher energy component has also been observed at TeV energies, with GRB\,190114C representing the first published detection of such photons from a GRB \citep{Acciari2019, MAGIC2019, Ajello2020}. These observations provided the first evidence of possible synchrotron self-Compton (SSC) radiation long predicted to play a role in the high-energy GRB afterglow emission. Subsequent reports of TeV emission with various levels of significance in GRB afterglows, GRB\,160821B \citep{Acciari2021}, GRB\,180720B \citep{Abdalla2019}, GRB\,190829A \citep{Abdalla2021}, GRB\,201015A \citep{gcn28659}, and GRB\,201216C \citep{Fukami2021} have shown that this Very High Energy (VHE) component may common in GRB emission.

Here we present \gbm observations of \grb, by far the brightest GRB ever detected by \gbm, and for which there is an abundance of photons to accommodate the most thorough and exhaustive investigation of its spectral and temporal properties. In Section\,\ref{sec:discovery} we discuss the timeline of discovery for \grb and Section\,\ref{sec:data} explains how to properly use the available \gbm data products, given the unique challenges of analyzing this event. Section\,\ref{sec:temporal_analysis} provides an overview of the temporal structure within the prompt emission followed by an analysis of the spectral evolution in Section\,\ref{sec:prompt_emission_phases}. The analysis results of the afterglow period are presented separately in Section\,\ref{sec:afterglow}. We then discuss the bulk emission properties and energetics associated with \grb in Section\,\ref{sec:lorentz_factor} and Section\,\ref{sec:energetics}. Finally, we provide an interpretation of our analysis in Section\,\ref{sec:interpretation} before summarizing of our results in Section\,\ref{sec:summary}.

\section{Discovery} \label{sec:discovery}

On 2022 October 9 at 13:16:59.99 UTC (\t0), the \gbm flight software triggered on \grb. The \gbm gamma-ray burst coordinates network (GCN)\footnote[1]{\url{https://gcn.gsfc.nasa.gov/gcn3_archive.html}\label{ss:gcn}} \enquote{Trigger Notice} was automatically distributed, but due to issues in the ground segment, further notices containing classification and localization information were not disseminated. At the time of the \gbm detection, the GRB was occulted by the Earth for the Neil Gehrels Swift Observatory (\swift). When the source position first became visible, \swift was transiting the South Atlantic Anomaly (SAA) and was unable to collect data, as reported by \cite{Williams2023}. At 14:10:17 UTC \grb was detected by the \swift Burst Alert Telescope (BAT) and observations were performed by the \swift X-ray Telescope (XRT) and Ultra-violet Optical Telescope (UVOT) instruments, yielding a best localization at RA(J2000) = 19\texp{h} 13\texp{m} 3.48\texp{s} ($288.26452\deg$), DEC(J2000) = +19$^{\deg}$ 46' 24.6" ($19.77350\deg$) with a 90\%-confidence error radius of 0.61" \citep{gcn32632}. Due to the particularly bright signal and a localization within the Galactic plane, the \swift team initially classified the event as having a Galactic origin.

The \swift GCN notices initiated the automated \lat analysis pipeline which searched and identified a bright high-energy signal within the \swift interval. This in turn initiated a deeper inspection of the \gbm data which appeared as one of the brightest transients ever discovered by \gbm. Private communication between the \gbm team, the \lat Collaboration, and the \swift team led to the conclusion that the location of all of these transients were consistent within the respective positional uncertainties of each instrument. An initial picture of identifying the same, incredibly bright GRB, with the prompt emission being detected by \fermi and the afterglow being detected by \swift an hour later, became the working hypothesis. GCN circulars reporting the brightest GRB ever seen were promptly sent by all parties involved \citep{gcn32635, gcn32636, gcn32637} encouraging follow-up observations from the community. Prompt signal association was later confirmed by the InterPlanetary Network \citep{gcn32641}.

These initial GCN circulars were followed by more than 100 additional circulars, and some Astronomer's Telegram notices (ATels)\footnote[2]{\url{https://astronomerstelegram.org/}\label{ss:atel}}, reporting detections across the electromagnetic spectrum and upper limits from non-electromagnetic messengers. Some highlights from the prompt emission include:
\bullets{
    The first detection of TeV energy photons during the prompt emission by LHAASO (up to 18\,TeV$\textrm{;}$ \citealt{gcn32677});
    A redshift of z = 0.151 reported by the VLT \citep{gcn32648};
    The identification of rings from dust echoes from the \swift-XRT \citep{atel15661};
    Polarization observations from IXPE \citep{gcn32690, gcn32754};
    Unsaturated observations from a Solar instrument (STIX) \citep{gcn32661};
    Observations from GRB CubeSats like GRBAlpha and SIRI-2 \citep{gcn32685, gcn32746};
    Non-detections of neutrinos in both IceCube and KM3NeT \citep{gcn32665, gcn32741};
    Measurable disturbances in Earth's ionosphere \citep{gcn32744, gcn32745}
}
The Hubble Space Telescope performed photometry \citep{gcn32921} and the James Webb Space Telescope performed spectroscopy \citep{gcn32821, Levan2023, Fulton2023} of the potential supernova component and emission was still detectable at radio frequencies up to 4 months later \citep{gcn33305}.

\section{Fermi Data} \label{sec:data}

\fermi includes two scientific instruments, \gbm and \lat. \gbm is a wide-field ($>$8 sr) survey instrument comprised of twelve sodium iodide (NaI) detectors and two bismuth germanate (BGO) detectors \citep{Meegan2009}. The NaI detectors cover the energy range from 8\,keV to 1000\,keV and are oriented in different directions around the spacecraft as to observe the entire unocculted sky. The two BGO detectors are on opposite sides of the spacecraft and cover an energy range from 200\,keV to 40\,MeV. The \lat is a pair-conversion telescope at the zenith of the spacecraft and is sensitive to gamma-ray energies from 20\,MeV to more than 300\,GeV \citep{Atwood2009}.

The highest resolution \gbm data product is the Time-Tagged Event (\tte) data, tagging individual photons to 2 microsecond temporal precision at the full 128 spectral channel resolution of \gbm. Continuous Spectroscopy (\cspec) data has the same full spectral resolution as \tte data but is a binned dataset with temporal resolution as fine as 1.024\,s in the interval following the trigger. Continuous Time (\ctime) data has 8 spectral channels with a temporal resolution of 64 milliseconds following a trigger. 

In this work we also use the \lat Low Energy (\lle) data in the 30\,MeV to 10\,GeV energy range \citep{Pelassa2010}. All \lat \lle data are used within time ranges specified in \citet{grb221009a_fermi_lat_collaboration_2023}. Both \gbm and \lle data are available for download via the public archive at the Fermi Science Support Center (FSSC) website\footnote[3]{\url{https://fermi.gsfc.nasa.gov/ssc/data/}}\texp{,}\footnote[4]{\url{https://heasarc.gsfc.nasa.gov/FTP/fermi/}}. For details about how to properly analyze \gbm data products, see Appendix\,\ref{app:fermi_intervals}. A complementary analysis including proper treatment of the \lat intervals of intense photon fluxes will be reported in \citet{grb221009a_fermi_lat_collaboration_2023}. 

\grb was detected by \gbm beginning at the trigger time (\t0) and lasting until \t0+1467\,s when it was occulted by the Earth. Prior to detection, the position of the GRB was within the field of view of \fermi beginning at \t0$-$2111\,s. \gbm detectors N3, N4, N6, N7, and N8 all had viewing angles within $60\deg$ of the burst at trigger time, but only detectors N4 and N8 stayed within $60\deg$ of the burst throughout the emission episode. These detectors, along with B1, are used throughout this analysis unless otherwise stated.

From the time of detection until Earth occultation, the incident photons hit \fermi at zenith angles ranging from 62$\deg$ to 110$\deg$ with respect to the \lat boresight and azimuth angles ranging from 258$\deg$ to 263$\deg$ (from 12$\deg$ to 7$\deg$ with respect to the BGO endcaps on the BGO detector plane). At such geometries, the incident photons travel through the BGO photomultiplier tubes and their housings, which are not adequately modeled in the detector response at low energies. We therefore omit the BGO data below 400\,keV throughout our analysis. The NaI spectra show deviations between the model and the data below 20\,keV that needs further analysis; these NaI data are therefore omitted from the spectral analysis.

Due to the extraordinarily high photon flux produced by \grb, both \gbm and \lat experienced periods with data issues\footnote[5]{\url{https://fermi.gsfc.nasa.gov/ssc/data/analysis/grb221009a.html}\label{ss:btis}} \citep{gcn32642, gcn32760, gcn32916}. We identify these periods as bad time intervals (BTIs). In \gbm, during these BTIs, the majority of \tte data are unrecoverable due to the summed count rate of all detectors exceeding the 375\,kHz data rate limit of the \gbm high speed science data bus, causing loss of \tte telemetry packets \citep{Meegan2009}. We therefore limit our analysis to the pre-binned 1.024\,s \cspec data within these regions, which are available without data loss, and use additional \lat \lle data, when available.

High counting rates create deadtime which is automatically corrected for in the exposure value in \gbm FITS files using Equation 4.24 in \citet{Knoll2010}. At input count rates above \abt50k counts per second (cps) more complex deadtime effects such as pulse pile-up (PPU) can occur \citep{Meegan2009}. PPU occurs when the overlap between electronic pulses in \gbm causes distortions in both the observed spectral shape and intensity \citep{Chaplin2013, Bhat2014}. This rate was significantly exceeded during the reported \gbm BTIs and cannot be corrected for automatically via the standard method. In order to obtain a full picture of \grb, we perform PPU-correction within the \gbm BTIs using the method described in Section\,\ref{subsec:BTIs}.


\subsection{Pulse Pile-up Correction} \label{subsec:BTIs}

A number of instruments have successfully applied PPU-correction techniques to a wide range of gamma-ray transients \citep{Mazets1999, Lysenko2019, Mailyan2016, Mailyan2019}. The effects of PPU on \gbm data and its correction have been analytically studied in the work of \cite{Chaplin2013} and correction techniques have been verified through Monte Carlo simulations and lab experiments with radioactive sources \citep{Bhat2014}. This analytical PPU method successfully achieved spectral fits of \gbm terrestrial gamma-ray flashes that could not have been satisfactorily fit without PPU corrections \citep{Mailyan2016, Mailyan2019}. In our work we apply the \gbm PPU-correction technique to the \gbm \cspec data for \grb.

The technique parallels that of a normal spectral analysis with a few additional steps to account for the distorted data. We first assume a parameterized photon model and forward fold it through the \gbm detector response matrix (DRM) to obtain a detector count spectrum. The count spectrum is then adjusted for PPU effects using the analytic method of \citet{Chaplin2013} and an assumed count rate. Since the analytical PPU-correction model includes all deadtime effects, the exposure time in the \gbm FITS files needs to be replaced with the observing time to avoid double counting deadtime. The PPU-adjusted count spectrum is then compared to the observed count spectrum. The assumed photon model parameters and count rate are then iteratively adjusted to maximize the likelihood between the PPU-adjusted counts spectrum and the observed counts spectrum. 

Unlike for the Cs$^{137}$ and Co$^{60}$ sources used when testing a BGO detector in a lab \citep{Bhat2014}, we do not know the true count rate or spectrum of \grb. Additionally, there are some residual uncertainties concerning the reliability of the PPU-correction technique at such extreme counting rates. Since our primary goal is to determine the energetics of this GRB we assume a simple Band function throughout the PPU reconstruction process and do not attempt to improve these fits by adding spectral breaks or additional spectral components. We do not take into account uncertainties associated with our non-optimal choice of spectral shape. For reasons which will be discussed in Section\,\ref{subsec:primary_pulse} and Section\,\ref{sec:energetics}, we consider this technique reliable for determining the total energetics of \grb and note that our results are consistent with those reported in \cite{Frederiks2023}, \cite{Ripa2023}, and \cite{An2023}.

\section{Temporal Analysis} \label{sec:temporal_analysis}

\begin{figure*}[h!tbp]
    \centering
    \includegraphics[width=\textwidth]{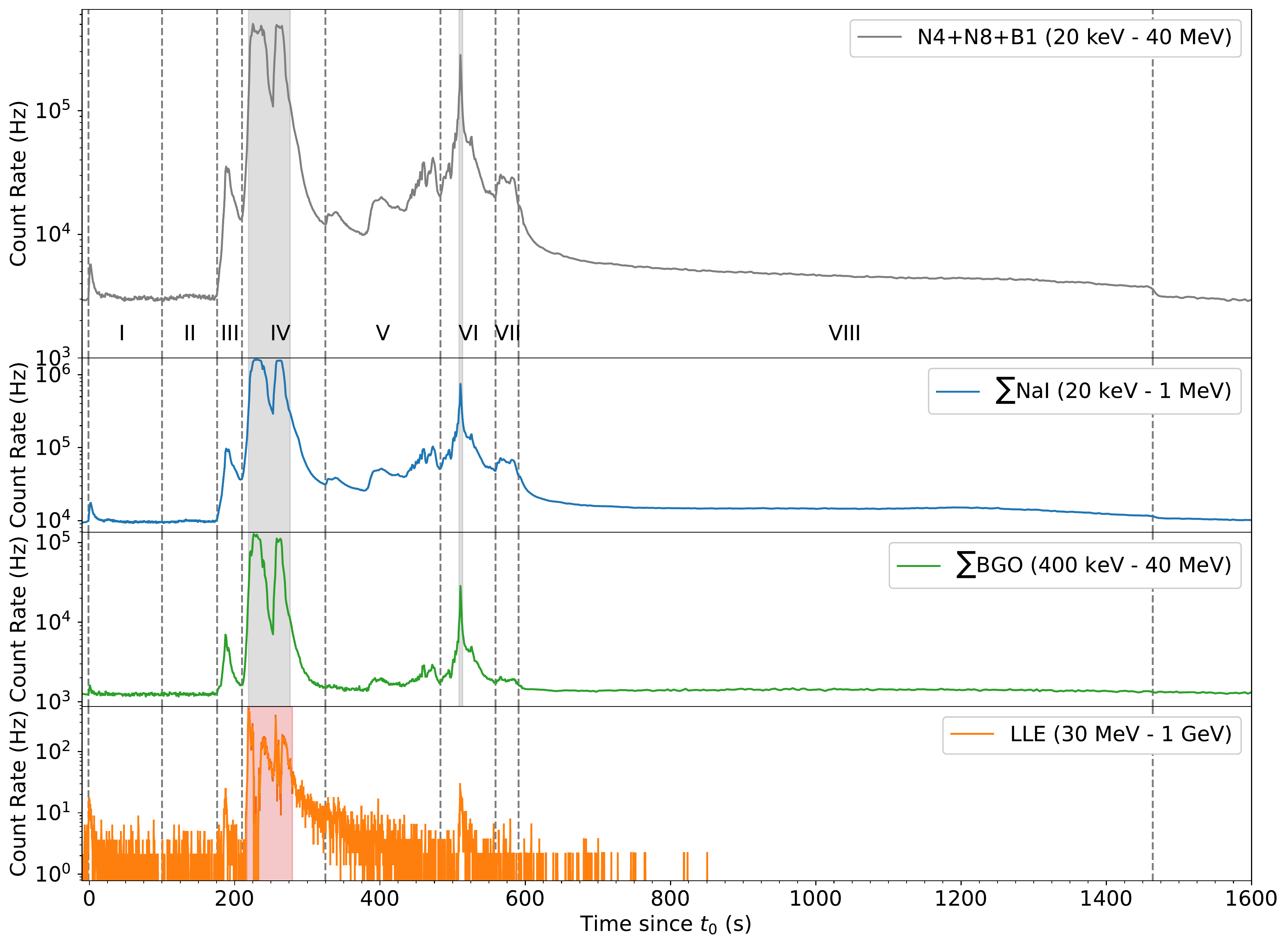}
    \includegraphics[width=\textwidth]{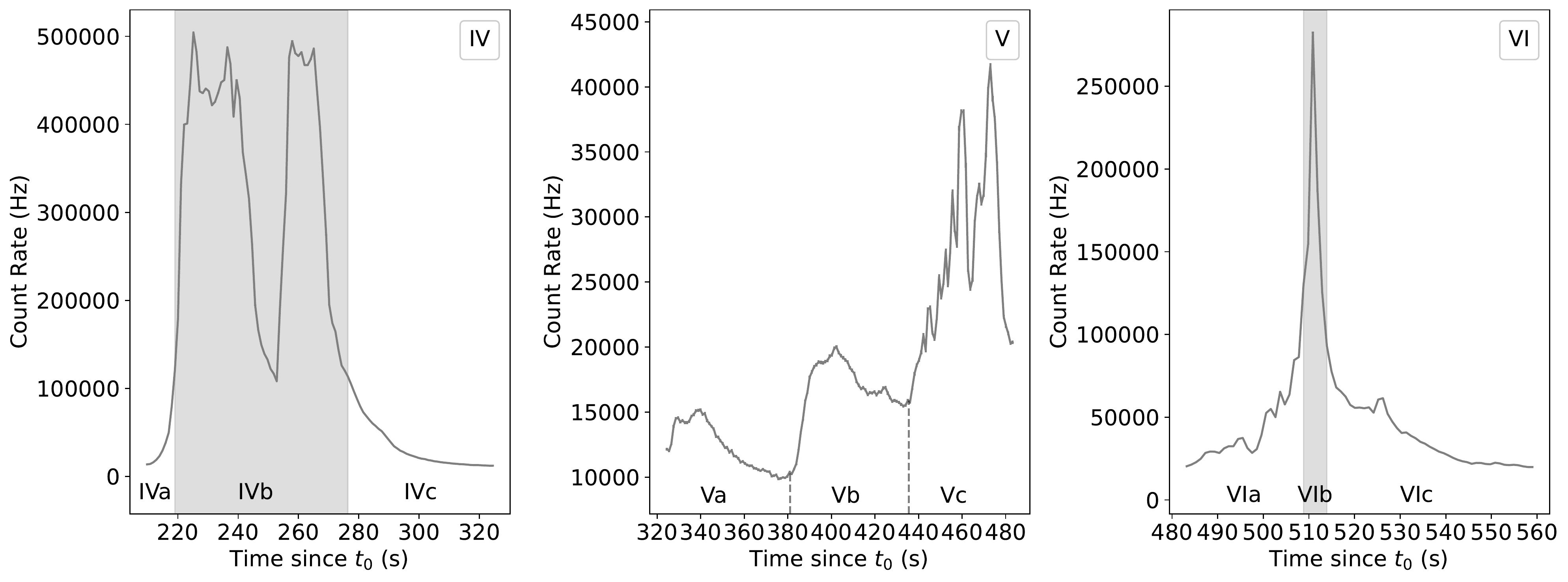}
    \caption{The top plot shows the uncorrected lightcurve of \grb in the 20\,keV to 40\,MeV energy range as seen by the three \gbm detectors with the lowest continuous viewing angles. The uncorrected lightcurve is divided into eight time intervals (I-VIII) differentiated by vertical dashed lines. Beyond region VIII (right of the last vertical dashed line at $t_{0}+1464$\,s) marks the time when \grb was occulted by the Earth. Intervals IV, V, and VI are further subdivided into three sub-intervals shown in the bottom three panels in the same energy range (20\,keV to 40\,MeV). The NaIs, BGOs, and \lle plots show the uncorrected lightcurve of \grb in different energy bands. The two gray vertical shaded regions in the \gbm plots denote the BTIs of \gbm (\t0+219.0-\t0+277.0\,s \& \t0+508.0-\t0+514.0\,s). The red vertical shaded region in the \lle plot denotes the revised BTI of \lat (\t0+217 to \t0+280\,s).}
    \label{fig:main}
\end{figure*}

Figure\,\ref{fig:main} shows the lightcurve of \grb as seen by \gbm and \lat with the \gbm BTIs highlighted with gray vertical shaded regions and the \lat BTI highlighted with a red vertical shaded region. Here we use the revised \lat BTI region, the details of which can be found in \citealt{grb221009a_fermi_lat_collaboration_2023}. Pulses are the basic unit of measure for GRB emission \citep{Hakkila2014}. Pulses are simple structures underlying the more complex GRB emission and are often superposed with one another. With this motivation, we separated the lightcurve of \grb into eight distinct emission intervals, and some sub-intervals, based on its morphology:
\begin{enumerate}[label=\Roman*)]
    \item The \enquote{Triggering Pulse} (\t0-1\,s -- \t0+43\,s)
    \item The \enquote{Quiet Period} (\t0+121\,s -- \t0+164\,s)
    \item The beginning of the primary emission which we refer to as the\enquote{Pre-Main Pulse} (\t0+177\,s -- \t0+210\,s)
    \item The \enquote{Primary Pulse} which contains the first \gbm BTI (\t0+210\,s -- \t0+324\,s)
    \begin{enumerate}[label=IV\alph*)]
        \item The interval before the first \gbm BTI
        \item The first \gbm BTI region
        \item The interval after the first \gbm BTI
    \end{enumerate}
    \item The smooth, then variable \enquote{Intra-Pulse Period} (\t0+326\,s -- \t0+483\,s)
    \item The \enquote{Secondary Pulse} which contains the second \gbm BTI (\t0+483\,s -- \t0+546\,s)
    \begin{enumerate}[label=VI\alph*)]
        \item The interval before the second \gbm BTI
        \item The second \gbm BTI region
        \item The interval after the second \gbm BTI
    \end{enumerate}
    \item The \enquote{Final Pulse} (\t0+546\,s -- \t0+597\,s)
    \item The \enquote{Afterglow} (\t0+597\,s -- \t0+1467\,s)
\end{enumerate}
Another motivation for these many intervals and sub-intervals is track the spectrotemporal evolution of this extremely bright GRB which is known to exist from previous GRB observations (e.g., \citealt{Liang1996}). These lightcurve intervals and their corresponding sub-intervals will be referenced throughout this work.


\subsection{Periodicity Searches} \label{subsec:qpo}

In light of the recent quasi-periodic oscillation (QPO) study of GRB\,211211A reported in \citet{Xiao2022} and of GRB\,910711 and GRB\,931101B reported in \citet{Chirenti2023}, we performed a QPO search on \grb using the high time-resolution \gbm \tte data across a wide range of frequencies using two different methods. 

We separately searched interval I (the triggering pulse), intervals III+IVa (the beginning of the main emission period before the first \gbm BTI), intervals IVc+V+VIa (the emission between the two \gbm BTIs), and intervals VIc+VII+VIII (the end of the main emission period after the second \gbm BTI). Since the loss of \tte packets during the \gbm BTIs creates artificial structure in the lightcurves, we do not consider these regions in our analysis. Visual inspection of sub-interval Vc suggests potentially interesting variability, thus we also search that segment separately.

We search for QPOs using Fourier-based methods (e.g., \citealt{vanderklis1989}) at frequencies $\gtrsim 20 \,\mathrm{Hz}$, where the overall variability of the GRB is relatively unimportant and photon counting noise dominates. We generate periodograms up to $5000\,\mathrm{Hz}$, and look at both linearly and logarithmically binned periodograms. As a threshold for significance, we choose $p < 0.001$ ($\sim3\sigma$), corrected for the number of frequencies searched. We find no credible QPO detection in any of the segments in the white noise-dominated regime above $20\,\mathrm{Hz}$. In all cases the significance remains far below the threshold corrected for the number of trials.

For frequencies $<20\,\mathrm{Hz}$, we follow the formalism in \citet{Huebner2022} and use a Bayes factor comparing a Gaussian process with a covariance function describing aperiodic red noise variability to a Gaussian process with a covariance function describing a combination of aperiodic red noise variability and a QPO. We modelled all segments individually with the Gaussian process, and refer the reader to \citet{Huebner2022} for more details on the analysis, including prior choices. We find no credible detections in intervals I, III+IVa, or VIc+VII+VIII, and focus our attention on region IVc+V+VIa and its sub-interval Vc, where the variability suggests the possible presence of a low-frequency signal. Indeed, in the model including both aperiodic variability and a QPO signal the posterior probability density for the QPO period is well-constrained to $12.34^{+2.12}_{-1.67}\,\mathrm{s}$. The Bayes factor for the two models is,

\begin{equation} \label{eq:bayes_factor}
    \log\p{\mathcal{B}} = \log\p{\frac{\mathcal{L}\p{D | M_1}}{\mathcal{L}\p{D | M_2}}} = 2.21
\end{equation}

\noindent
where $M_1$ is the model with QPO, $M_2$ is the model without QPO, and $D$ is a placeholder for the data, in this case the lightcurve. A Bayes factor of $2.21$ is considered moderate evidence for a QPO, but we caution the reader that Bayes factors are extremely sensitive to prior choices. A more detailed QPO analysis will be deferred to a future study.


\subsection{Minimum Variability Timescale} \label{subsec:mvt}

GRB lightcurves exhibit variability on various time scales. Internal shocks can produce the observed complex temporal structure provided that the source itself is variable. The minimum variability timescale (MVT; $\Delta t_{\textrm{min}}$) is a measure of the shortest coherent variations of the lightcurve \citep{Bhat2012} and, in turn, can be related to the minimum Lorentz factor of the relativistic shells emitted by the central engine. This measure carries information about the variability timescale of the central engine itself. 

Here we use the statistical method described in \citet{Bhat2013} to estimate the MVTs of a GRB using \gbm \tte data. We first identify the prompt emission region and an equal duration of background region. We then derive a differential of each lightcurve and compute the ratio of the variances of the GRB prompt region to that of the background region. This is repeated by varying the bin widths of the lightcurve starting at sub-millisecond values. This ratio, divided by the bin width, is plotted as a function of bin width. At very fine bin widths this ratio falls monotonically with increasing bin width (1/bin width variation) signifying that at fine bin widths the variations in the background and burst regions are statistically identical. In other words, the signal in the burst lightcurve is indistinguishable from Poissonian fluctuations. At some point, as the bin widths increase, the ratio starts increasing with bin width. The point where the correlation between the ratio and bin width change is defined as the minimum variability timescale. We measure the bin width at this valley by fitting it with a quadratic function with the $\Delta t_{\textrm{min}}$ being the minimum of the quadratic \citep{Bhat2013}.

\begin{figure*}[h!tbp]
    \centering
    \includegraphics[width=\textwidth]{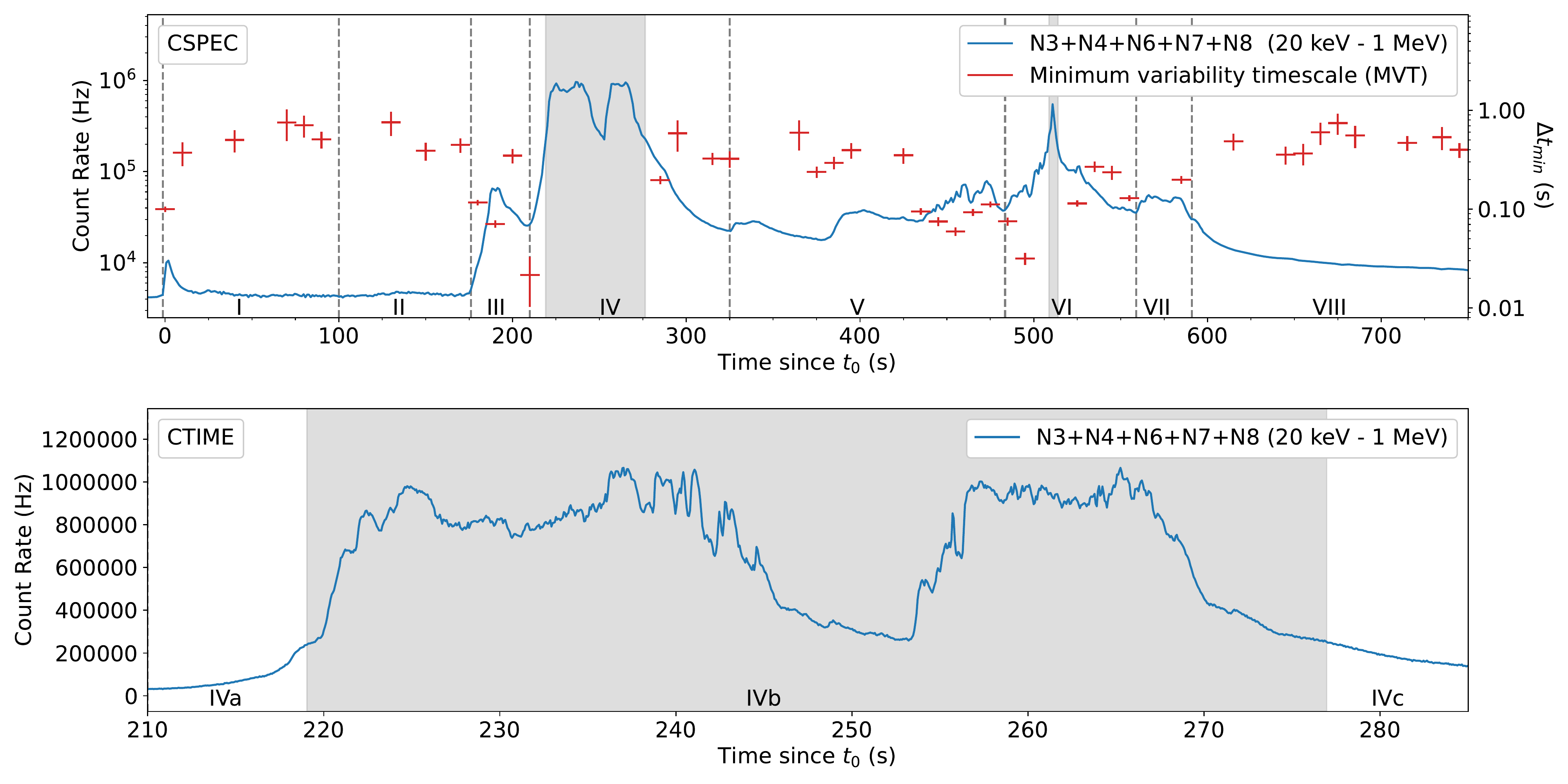}
    \caption{Top: Uncorrected \gbm \cspec data (1.024\,s resolution) in the five NaI detectors with viewing angles to \grb below 60$^{\deg}$ at trigger time and the evolution of the minimum variabtility timescale (MVT; $\Delta t_{\textrm{min}}$) obtained from the \gbm non-BTI regions. The MVT ranges from $\sim$0.01\,s to $\sim$1\,s. The decreasing MVT occurs in regions where the count rate is more variable. Bottom: Uncorrected \gbm \ctime data (0.064\,s resolution) for the same two NaI detectors during the first \gbm BTI region.}
    \label{fig:mvt}
\end{figure*}

Usually, a single MVT value is quoted for each burst. However, in the case of \grb, it is possible to derive $\Delta t_{\textrm{min}}$ as a function of time. Figure\,\ref{fig:mvt} shows the evolution of $\Delta t_{\textrm{min}}$ together with the \gbm NaI lightcurve (8 - 1000\,keV). The $\Delta t_{\textrm{min}}$ has been shown to be equal to the shortest rise of a lognormal pulse if one tries to fit the entire lightcurve as a superposition of multiple lognormal pulses \citep{Bhat2013}. With low counting statistics, one would expect the number of detected photons within a certain time bin to anti-correlate with $\Delta t_{\textrm{min}}$, as has been seen in previous GRB observations. However, the MVT is not a measure of the count rate, but rather a measure of how fast the count rate is changing. Therefore, Figure\,\ref{fig:mvt} does not show an anti-correlation between the the count rate and $\Delta t_{\textrm{min}}$, but rather between the variability of the count rate and $\Delta t_{\textrm{min}}$. Meaning $\Delta t_{\textrm{min}}$ is not determined by Poissonian noise, but instead by the intrinsic time variability thanks to the exceptional brightness of this GRB.

During the period affected by PPU, we use the \gbm \ctime data with 64\,ms resolution to visually inspect the lightcurve. We clearly identify pulses that have a rise time of one temporal bin. Based on this observation, and the fact that there is an anti-correlation between the change in count rate and the MVT, we conservatively assume a variability timescale of 0.1\,s during the PPU period.

\subsection{Duration} \label{subsec:duration}

The standard duration measure of a GRB ($\textrm{T}_{90}$) is defined as the temporal interval containing 90\% of the total time-integrated photon flux in the 50–300\,keV energy range. The standard $\textrm{T}_{90}$ analysis using the \gbm \ctime files without PPU-correction overestimates the duration of \grb, as the count rate of the brightest portions of the lightcurve are decreased in intensity due to PPU and deadtime effects. To mitigate these effects, we used the PPU-corrected \cspec data within the two \gbm BTIs. Because a Band function was assumed in our PPU-correction method (see Section\,\ref{subsec:BTIs}), we performed time-integrated fits with a Band function to each of the non-BTI sub-interval shown in Figure\,\ref{fig:main}. We then integrated each fit from 50-300\,keV to obtain a PPU-corrected $\textrm{T}_{50}$ and $\textrm{T}_{90}$ measurement. For \grb, we estimate $\textrm{T}_{50}=25.6\pm1.0$\,s and $\textrm{T}_{90}=289\pm1$\,s. The duration $\textrm{T}_{90}=289$\,s is longer than 97.5\,\% of all \gbm GRBs and longer than 96.4\,\% of all long \gbm GRBs based on the classification of \citet{vonKienlin2020}.

\section{Prompt Emission Phases} \label{sec:prompt_emission_phases}

\begin{figure*}[h!tbp]
    \centering
    \includegraphics[width=\textwidth]{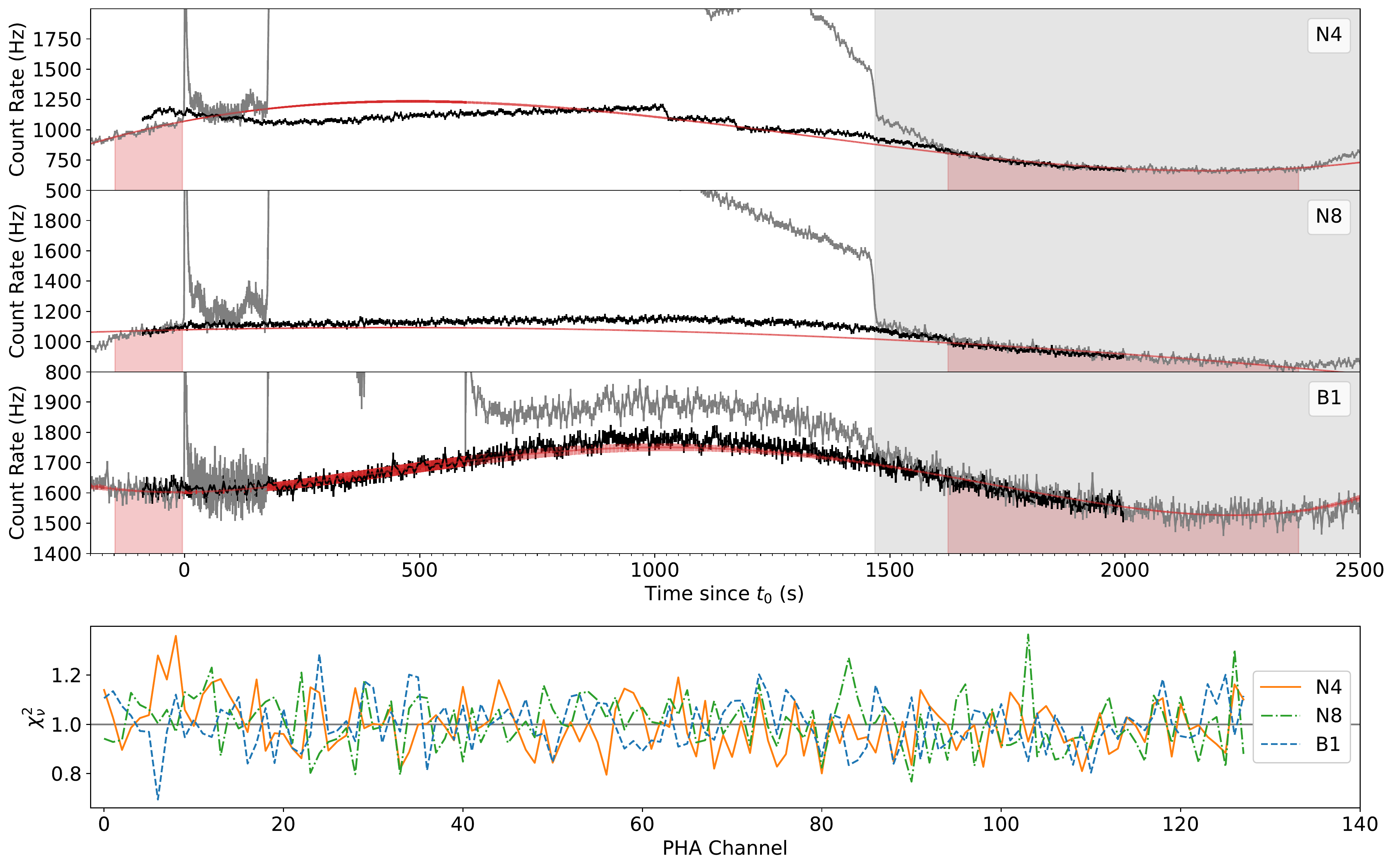}
    \caption{Top: The gray lightcurves are of \grb as seen in \gbm detectors N4, N8, and B1. The black lightcurves are the averaged background from 30 orbits before and 30 orbits after the \gbm trigger time, when \fermi was in the same orbital location and orientation. The red shaded regions mark the background selection regions used for polynomial fitting. The red lines are the 4\texp{th} order polynomial fits to the \grb lightcurves that best match the averaged orbital background. The gray shaded area on the right marks the time when the GRB was occulted by the Earth. The discrepancy at $t=0$ between the gray GRB lightcurve and the black averaged background lightcurve in panel N4 is due to the propagation of the Earth occultation steps in the orbital fits (as seen around 1100\,s and 1200\,s). The polynomial order and background range can be found in Section\,\ref{sec:prompt_emission_phases}. Bottom: The $\chi^{2}_{\nu}$ fits of the polynomial fit to the \grb lightcurve in each of the 128 \cspec energy bins for each detector.}
    \label{fig:background_osv}
\end{figure*}

Throughout our spectral analyses, we test a variety of standard spectral models including a power-law (PL), a power-law with an exponential cutoff (COMP), a Band function (Band; \citealt{Band1993}), a blackbody (BB), and combinations thereof \citep{Goldstein2012, Gruber2014, Poolakkil2021}. We additionally investigate two more complex spectral models, a double smoothly broken power-law (DSBPL; \citealt{Ravasio2018}) and a multicolor blackbody (mBB; \citealt{Hou2018}). When performing these spectral analyses, we used the Rmfit version 4.3.2\textsuperscript{\ref{ss:gbmsoftware}} spectral fitting package and the \gbm Data Tools\textsuperscript{\ref{ss:gbmsoftware}} Python software package \citep{GbmDataTools}. For analyses that could be performed by both programs, results were compared and shown to be consistent. 

The desire to investigate more complex spectral models (e.g., DSBPL and mBB) for \grb resulted in the extension of the \gbm Data Tools functionality to include a new Bayesian Markov Chain fitting technique, similar to the \textit{McSpecFit} package described in \cite{Zhang2016}. The validity of our Bayesian Markov Chain fitting technique was tested against the traditional \gbm Data Tools and Rmfit fitting routines using the standard spectral models mentioned previously and was shown to produce consistent results.

Figure\,\ref{fig:background_osv} shows that the prompt emission of \grb remained above background for most of its duration in \gbm. The \abt1200\,s of continuous emission means the standard polynomial background estimation method has a higher chance of misrepresenting the true background rate. To mitigate this, we used the orbital background estimation method\footnote[6]{\url{https://fermi.gsfc.nasa.gov/ssc/data/analysis/user/}\label{ss:gbmsoftware_osv}} for \gbm data presented in \citet{Fitzpatrick2011} which allowed us to better understand the background trend throughout the prompt emission phase. We performed a standard polynomial fitting technique to the lightcurve over a known background region on either side of the emission episode while adjusting the polynomial fit to best match the orbital background estimate (see Figure\,\ref{fig:background_osv}). The $\chi^{2}_{\nu}$ fit statistic for our polynomial fitting technique across all 128 \cspec energy channels is shown at the bottom of Figure\,\ref{fig:background_osv}. For the \lle data we used the standard polynomial fitting method with polynomial order 2 and background selection regions from \t0-200\,s to \t0-2\,s and from \t0+550\,s to \t0+560\,s. Due to the exceptionally long duration of this event, multiple detector response matrices DRMs for intervals spanning the burst duration were generated via the \gbm Response Generator\footnote[7]{\url{https://fermi.gsfc.nasa.gov/ssc/data/analysis/gbm/}\label{ss:gbmsoftware}}. Both the polynomial background estimate and these DRMs were used throughout our spectral analyses.

All spectral fits discussed in this section correspond to the lightcurve intervals and sub-intervals shown in Figure\,\ref{fig:main} and their fit parameters can be found in Table\,\ref{tab:time-res_spec_fits}. The fit statistics quoted in Table\,\ref{tab:time-res_spec_fits} are higher than expected. We attribute this to our coarse time resolution of our spectral fits. The time intervals used are likely not well fit because the spectral evolution evolves on a very short time scale and produces time-integrated shapes that significantly deviate from the time-resolved shape.


\subsection{Pre-Trigger Interval} \label{subsec:pre-trigger_interval}

Using the \swift-XRT localization of \grb and \fermi's orbital location, we find the GRB was first visible to \gbm \abt2111\,s before \t0. We used the \gbm \targeted, an offline search algorithm designed to find subthreshold short GRBs in \gbm data, to search for emission prior to the \gbm trigger time \citep{Blackburn2015, Goldstein2016}. No such emission was found. Using the \targeted spectral templates representative of spectrally-hard and \enquote{normal} GRBs, we set 3$\sigma$ flux upper limits over this pre-trigger interval at $5.1 \times 10^{-8}$ erg cm$^{-2}$ s$^{-1}$ and $8.0 \times 10^{-8}$ erg cm$^{-2}$ s$^{-1}$, respectively, for the 1-s timescale in the 10\,keV to 1\,MeV energy range. Using the redshift of \grb, we limit the isotropic luminosity of pre-trigger emission to $L_{iso} < 7.1 \times 10^{48}$ erg s$^{-1}$ and $L_{iso} < 7.5 \times 10^{48}$ erg s$^{-1}$, respectively.

\begin{figure*}[h!tbp]
    \centering
    \includegraphics[width=\textwidth]{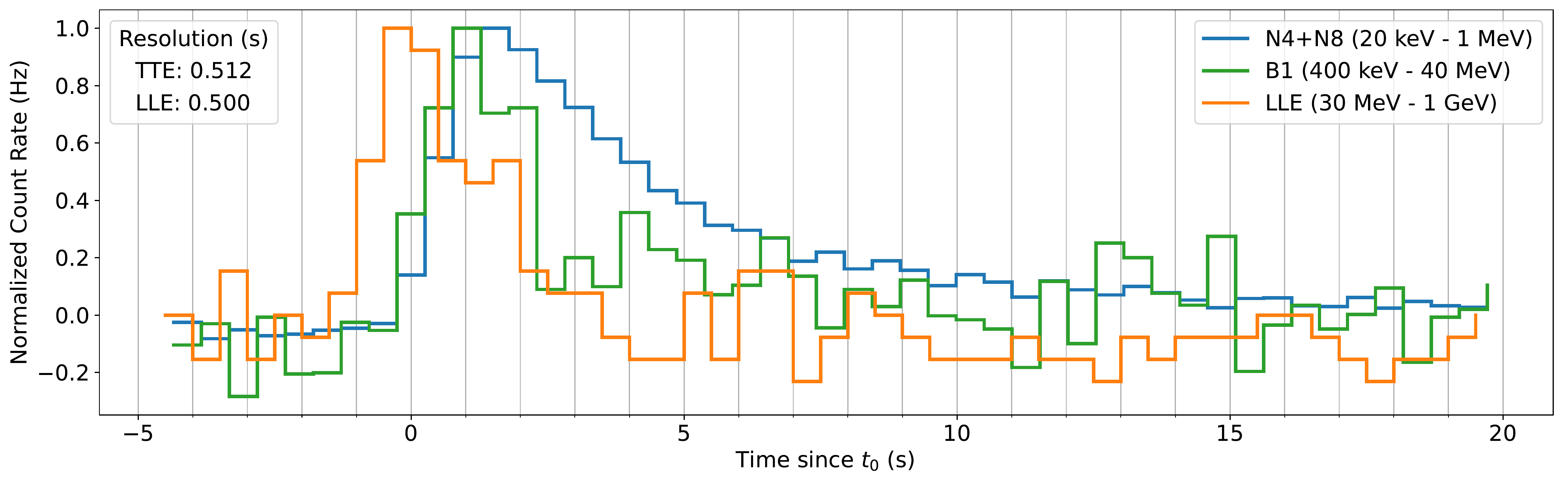}
    \begin{minipage}[b]{0.32\linewidth}
        \includegraphics[width=\textwidth]{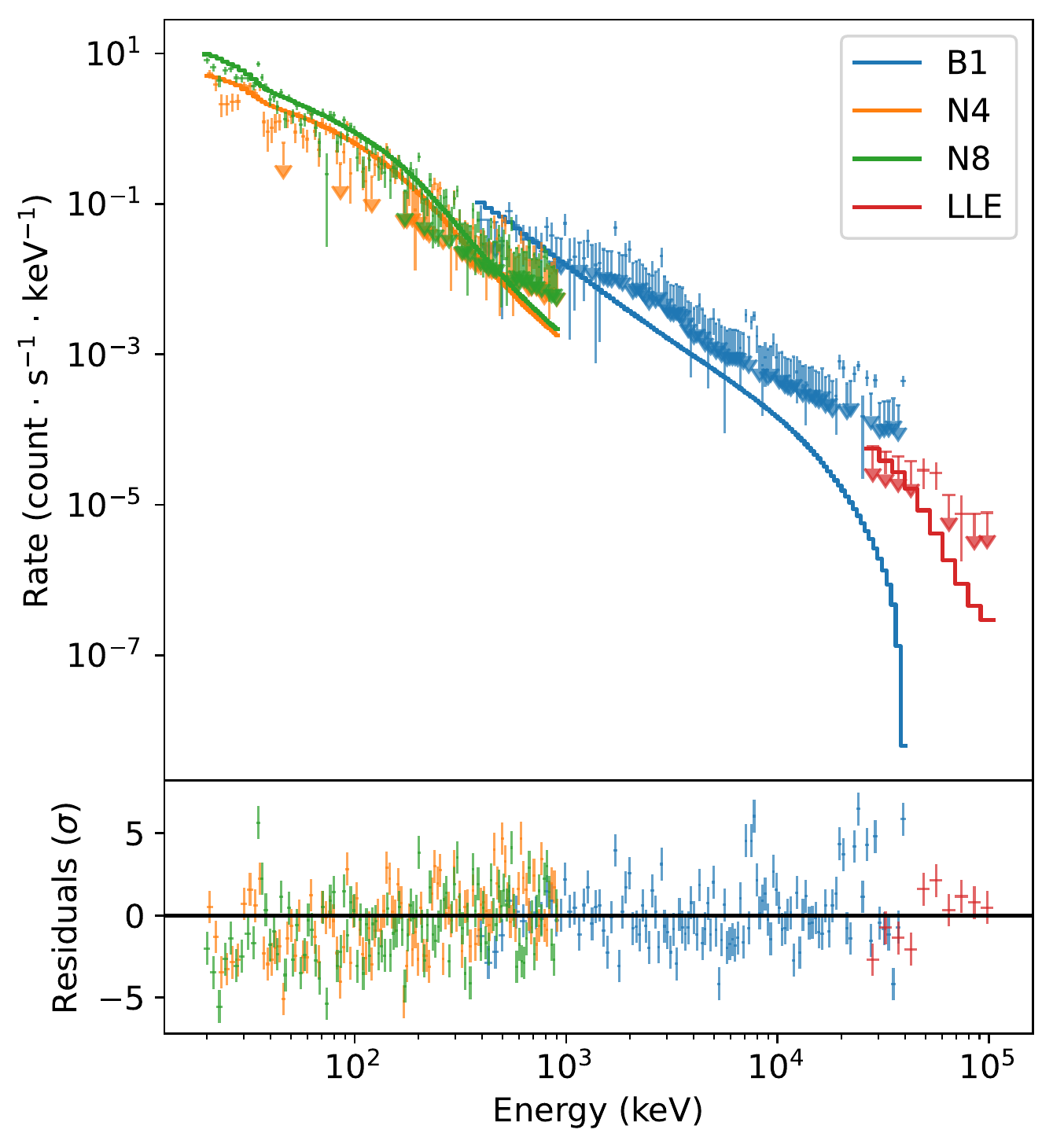}
    \end{minipage}
	\begin{minipage}[b]{0.32\linewidth}
        \includegraphics[width=\textwidth]{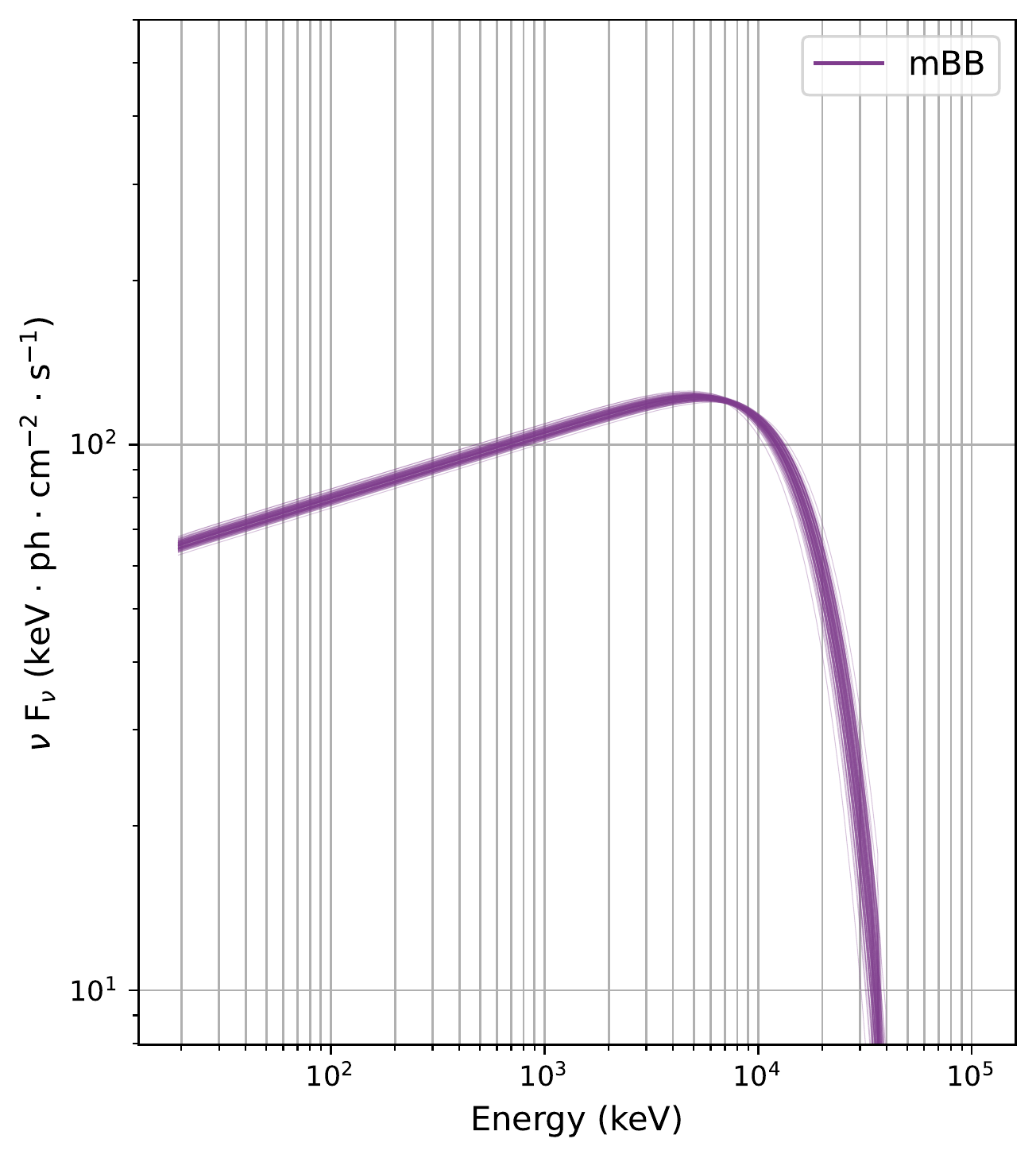}
    \end{minipage}
	\begin{minipage}[b]{0.33\linewidth}
        \includegraphics[width=\textwidth]{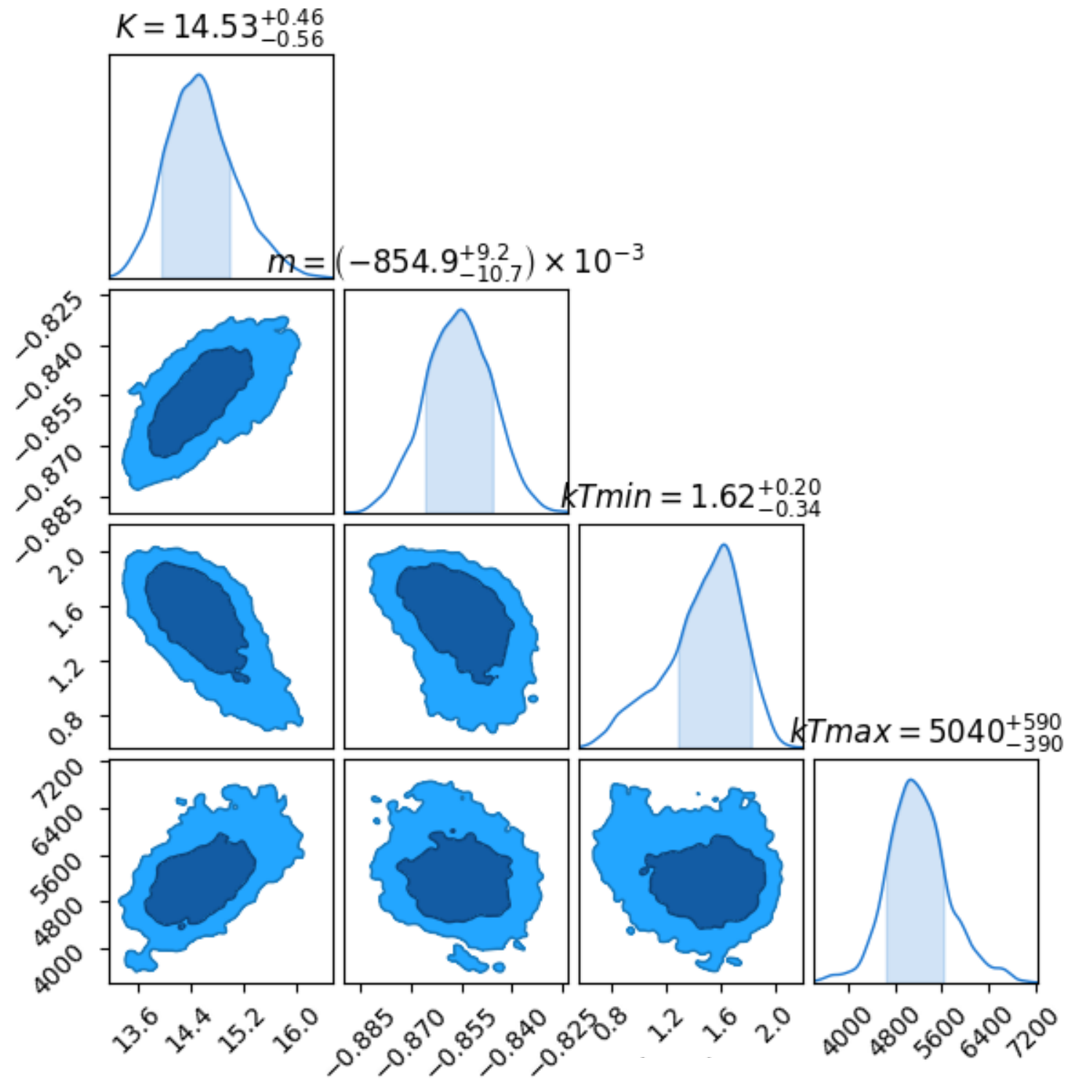}
    \end{minipage}
    \caption{Top: Background subtracted and normalized lightcurves of 0.512\,s binned \gbm \tte data and 1.0\,s binned \lle data for the triggering pulse with the higher-energy photons arriving prior to the lower-energy photons. Bottom: The fitted counts spectrum (left), the model spectrum (center), and the posteriors of the model parameters (right) for the multicolor blackbody function fitted to the triggering pulse.}
    \label{fig:precursor}
\end{figure*}


\subsection{Triggering Pulse} \label{subsec:triggering_pulse}

The triggering pulse (Figure\,\ref{fig:main}, region I) was analyzed using a combination of \gbm \cspec and \lat \lle data. Figure\,\ref{fig:precursor} shows high-energy \lle photons arriving \abt1.5\,s earlier than lower-energy BGO and NaI photons, with a lag that increases as energy decreases. Although distinct, this behavior is not unique as it has been previously observed in the initial pulse of GRB\,130427A as well \citep{Ackermann2014, Preece1998}. 

The first interval, referred to as sub-region Ia in Table\,\ref{tab:time-res_spec_fits}, spans the first 8 seconds of region I and was best fit by a COMP function with low-energy photon index $\alpha=-1.69$ and $\textrm{E}_{peak}=4$\,MeV. We compared these values to the corresponding time-integrated parameter distributions in the \gbm 10-year Spectral Catalog \citep{Poolakkil2021} and find that only 0.26\% of bursts (6/2295) have $\alpha<-1.69$ and 0.23\% have $\textrm{E}_{peak}$ above 4\,MeV (3/1311). While we note that this is not a direct comparison as this fit is neither time-integrated for a whole burst nor a 1.024\,s peak-flux interval, the $\alpha$ and $\textrm{E}_{peak}$ values are both outliers at $\sim$3$\sigma$ significance.

The full triggering pulse (region I), from \t0-1.3\,s to \t0+42.9\,s, was best fit with a COMP model (see Table\,\ref{tab:time-res_spec_fits}). The mBB model, comprised of a superposition of blackbodies with different temperatures, can produce the COMP model when the power-law index (m) equals zero \citep{Hou2018}. We tested whether the triggering pulse of \grb is consistent with a quasi-thermal origin by fitting region I with this function. Although the best fit does not reproduce the COMP spectra, these two models both produce adequate fits to the data. As shown in Table\,\ref{tab:time-res_spec_fits}, this fit yields a \kTmin of \abt1.6\,keV, and a \kTmax of \abt5\,MeV. Both counts and model spectra for this fit are shown in Figure\,\ref{fig:precursor}. \kTmin and \kTmax can be converted into peak (or \enquote{break}) energies with the relation $E_{b,1}\sim3kT_{min}$ (or $E_{b,2}\sim3kT_{max}$) \citep{Hou2018}. This yields $E_{b,1}\approx4.8$\,keV and $E_{b,2}\approx15$\,MeV. The \gbm+\lle energy range has lower and upper limits of \abt20\,keV and \abt300\,MeV, respectively for our analysis of \grb. This means we were only able to confidently constrain $E_{b,2}$ due to the addition of \lle data. The $E_{b,1}$ value from our analysis only tells us that the power-law index in the mBB fit extends below the \gbm energy range. The mBB model from \citealt{Hou2018} also gives a direct measure of luminosity. The fit value of $14.5_{-0.6}^{+0.5}$ with the scale $L_{39}/D^2_{L,10 \rm{kpc}}$ gives a corresponding value of $7.5_{-0.3}^{+0.3} \times 10^{49}$\,erg\,s$^{-1}$ at z=0.151 ($D_L = 724$\,Mpc).


\subsection{Quiet Period} \label{subsec:quiet}

We define the quiet period (Figure\,\ref{fig:main}, interval II) as the time between the triggering pulse and the onset of the main prompt emission phase. This region appears to have no emission, but careful analysis reveals detectable low-level emission during most of this time (see NaI panels of Figure\,\ref{fig:background_osv}). We associate the flux to \grb through consistent localizations of the low-level flux to the location of the GRB. The \t0+121\,s to \t0+164\,s interval, just before the onset of the main prompt emission, is best fit by a COMP function with an additional BB 

\begin{figure*}[h!tbp]
    \centering
    \begin{minipage}[b]{0.32\linewidth}
    	\includegraphics[width=\textwidth]{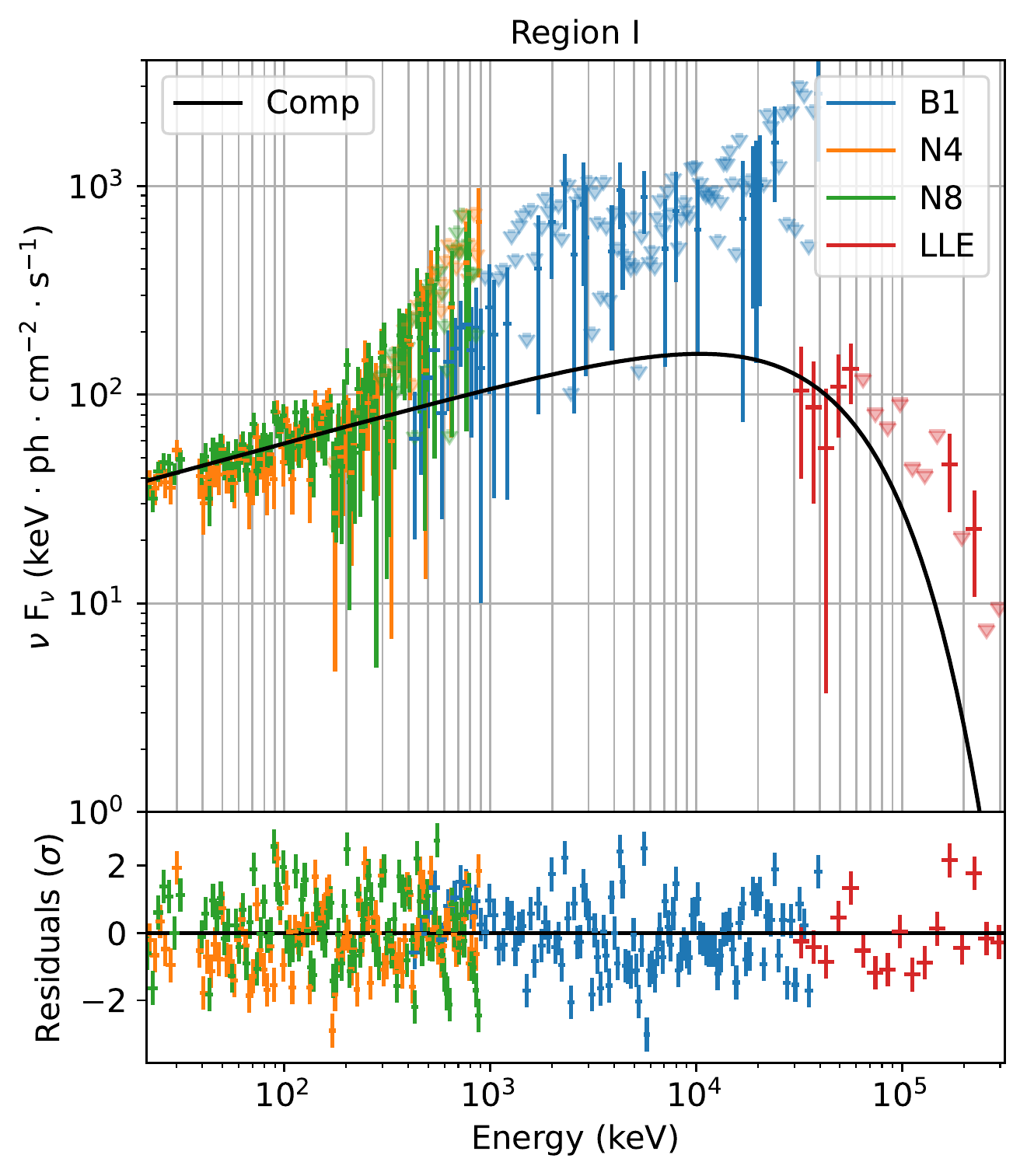}
    \end{minipage}
	\begin{minipage}[b]{0.32\linewidth}
        \includegraphics[width=\textwidth]{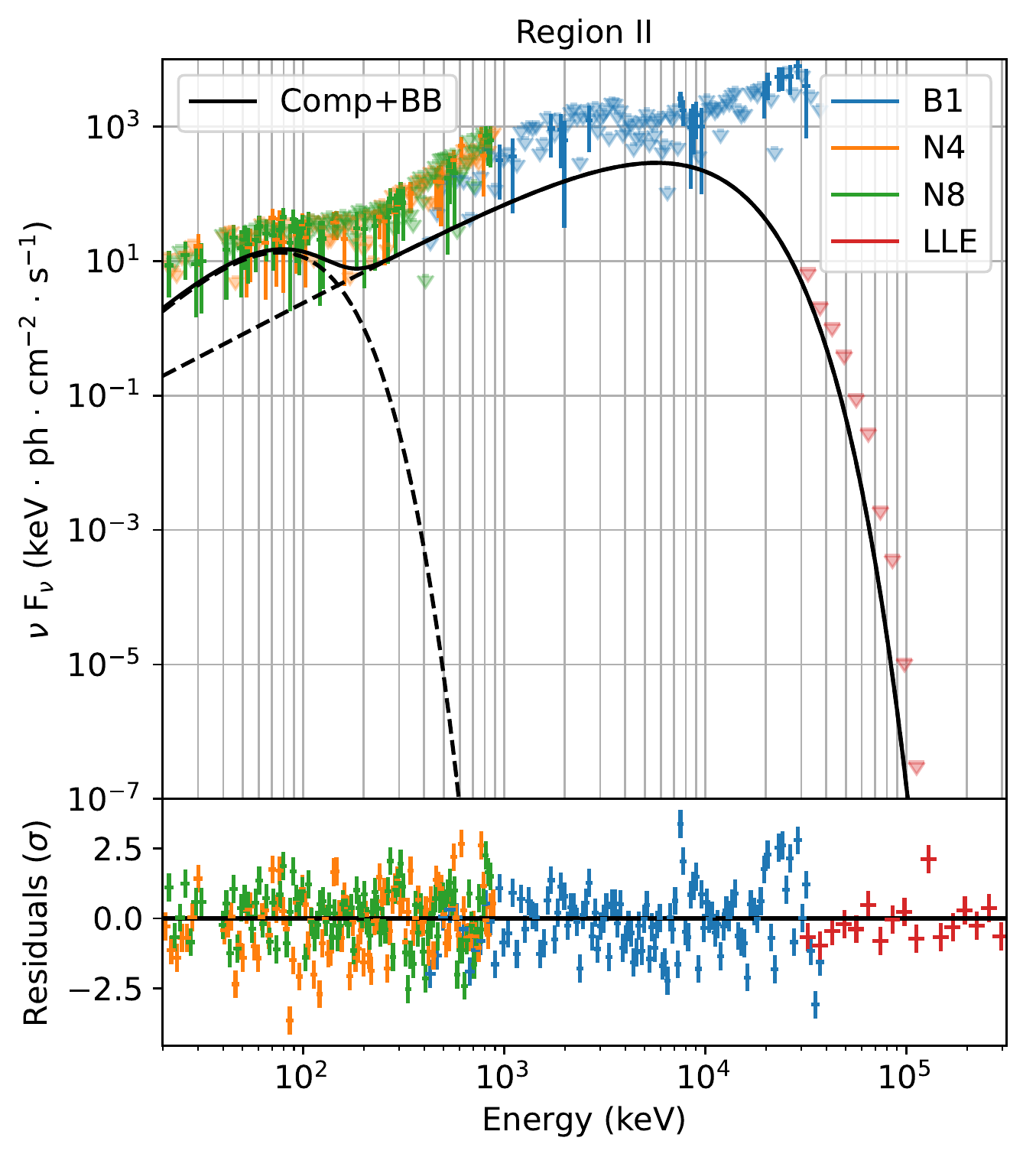}
    \end{minipage}
	\begin{minipage}[b]{0.32\linewidth}
        \includegraphics[width=\textwidth]{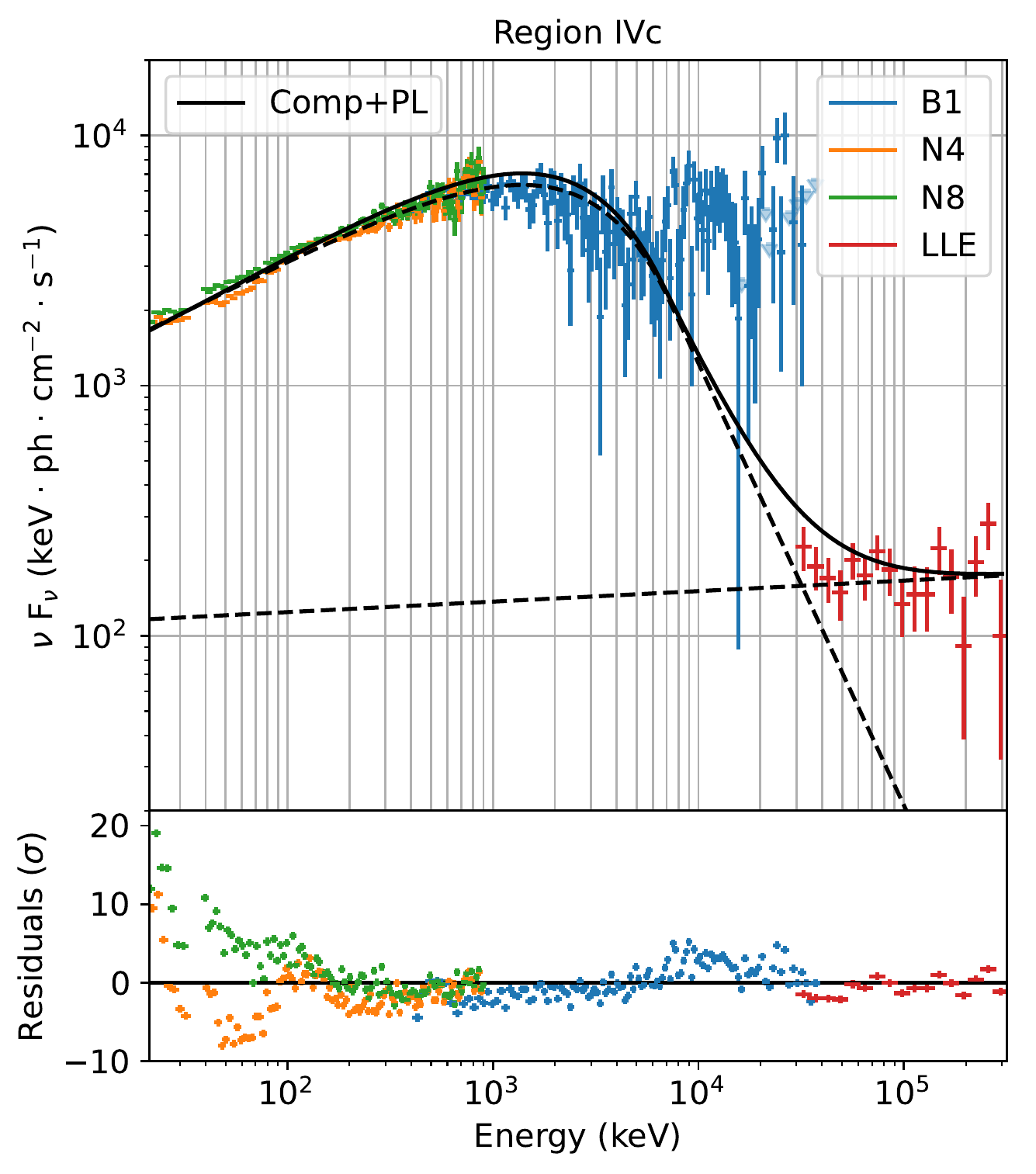}
    \end{minipage}
    \begin{minipage}[b]{0.32\linewidth}
        \includegraphics[width=\textwidth]{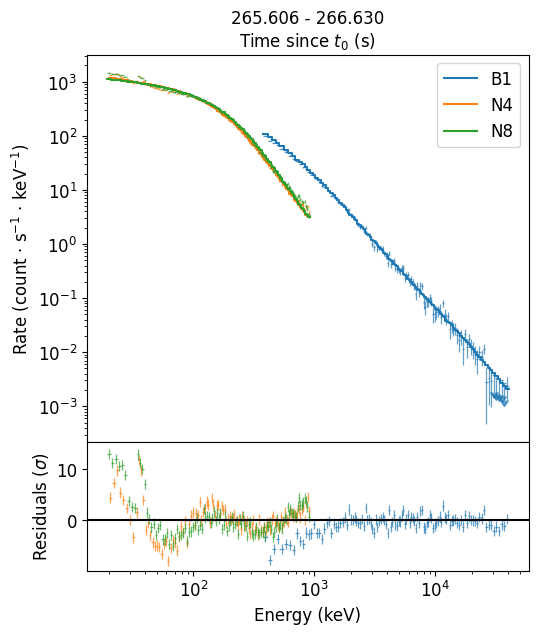}
    \end{minipage}
    \begin{minipage}[b]{0.32\linewidth}
        \includegraphics[width=\textwidth]{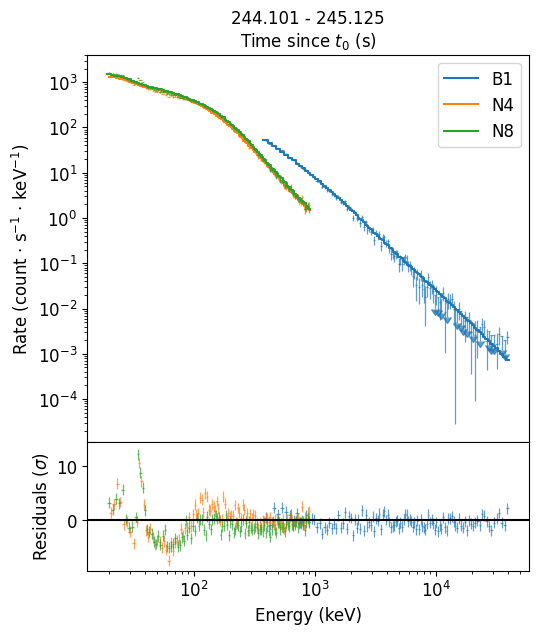}
    \end{minipage}
	\begin{minipage}[b]{0.32\linewidth}
        \includegraphics[width=\textwidth]{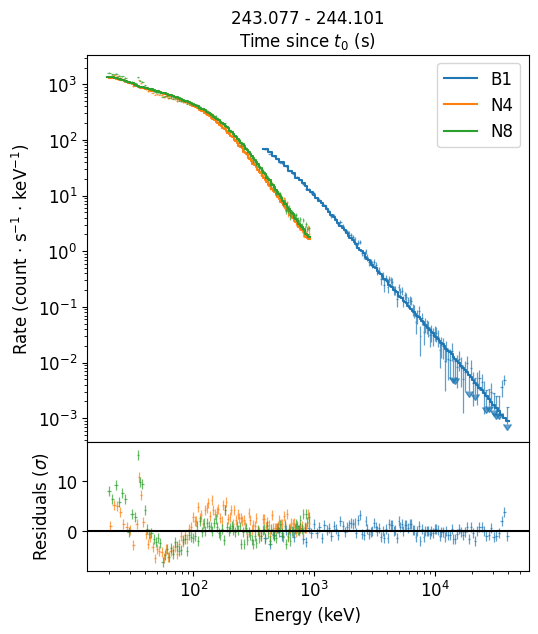}
    \end{minipage}
    \caption{Top: The $\nu$F($\nu$) spectra for region I best fit by a COMP (left), the $\nu$F($\nu$) spectra for region II best fit by a COMP+BB (middle) and the $\nu$F($\nu$) spectra for sub-region IVc best fit by a Band+PL (right). Bottom: Average examples of counts spectrum plots during the \gbm BTI region of sub-interval IVb (Left: \t0+243.1\,s, Middle: \t0+244.1\,s, Right: \t0+265.6\,s) achieved by using the PPU-correction method described in Section\,\ref{subsec:BTIs}. The spike at 511\,keV is ignored during the fitting process and does not affect the fit statistic.}
    \label{fig:ppu}
\end{figure*}

\begin{longtable*}{cc|ccccc|cc}
\label{tab:time-res_spec_fits}\\
\hline
\hline
\textbf{Model} & \textbf{Time} &  &  & \textbf{Model} &  &  & \multirow{2}{*}{\textbf{$\textrm{C}_{stat}$/DoF}} & \lle \\
\textbf{(region)} & \textbf{Range} &  &  & \textbf{Components} &  &  &  & \textbf{Used?} \\
\endfirsthead
\endhead
\hline
\hline
COMP & -0.003 &  &  & $\alpha$ & $\textrm{E}_{peak}$ &  & \multirow{2}{*}{485/343} & \multirow{2}{*}{Y} \\
\cline{3-7}
(Ia) & 8.576 &  &  & $-1.69\pm0.01$ & $3980\pm366$ &  &  &  \\
\hline
\hline
COMP & -1.343 &  &  & $\alpha$ & $\textrm{E}_{peak}$ &  & \multirow{2}{*}{497/340} & \multirow{2}{*}{Y} \\
\cline{3-7}
(I) & 42.881 &  &  & $-1.73\pm0.02$ & $10440\pm1900$ &  &  &  \\
\hline
mBB & -1.343 & K & $\textrm{kT}_{min}$ & m & $\textrm{kT}_{max}$ &  & \multirow{2}{*}{487/379} & \multirow{2}{*}{Y} \\
\cline{3-7}
(I) & 42.881 & $14.5^{+0.5}_{-0.6}$ & $1.62^{+0.20}_{-0.34}$ & $-0.854^{+0.009}_{-0.011}$ & $5000^{+600}_{-400}$ &  &  &  \\
\hline
\hline
BB+COMP & 121.219 &  & kT & $\alpha$ & $\textrm{E}_{peak}$ &  & \multirow{2}{*}{2169/334} & \multirow{2}{*}{Y} \\
\cline{3-7}
(II) & 164.228 &  & $19.27\pm1.66$ & $-0.4\pm0.2$ & $5722\pm564$ &  &  &  \\
\hline
Band & 176.516 &  &  & $\alpha$ & $\textrm{E}_{peak}$ & $\beta$ & \multirow{2}{*}{2130/338} & \multirow{2}{*}{Y} \\
\cline{3-7}
(III) & 210.309 &  &  & $-1.158\pm0.003$ & $1023\pm12$ & $-3.28\pm0.04$ &  &  \\
\hline
Band & 210.309 &  &  & $\alpha$ & $\textrm{E}_{peak}$ & $\beta$ & \multirow{2}{*}{1741/352} & \multirow{2}{*}{Y$^{*}$} \\
\cline{3-7}
(IVa) & 219.525 &  &  & $-1.159\pm0.003$ & $3664\pm47$ & $-2.70\pm0.03$ &  &  \\
\hline
Band+PL & 277.894 &  & Index & $\alpha$ & $\textrm{E}_{peak}$ & $\beta$ & \multirow{2}{*}{5707/336} & \multirow{2}{*}{Y$^{*}$} \\
\cline{3-7}
(IVc) & 323.975 &  & $-1.916\pm0.009$ & $-1.583\pm0.001$ & $1387\pm9$ & $-3.77\pm0.01$ &  &  \\
\hline
Band & 326.023 &  &  & $\alpha$ & $\textrm{E}_{peak}$ & $\beta$ & \multirow{2}{*}{3801/335} & \multirow{2}{*}{Y} \\
\cline{3-7}
(Va) & 381.023 &  &  & $-1.80\pm0.01$ & $69\pm1$ & $-2.24\pm0.01$ &  &  \\
\hline
Band & 381.023 &  &  & $\alpha$ & $\textrm{E}_{peak}$ & $\beta$ & \multirow{2}{*}{4217/335} & \multirow{2}{*}{Y} \\
\cline{3-7}
(Vb) & 435.546 &  &  & $-1.658\pm0.003$ & $520\pm11$ & $-2.98\pm0.02$ &  &  \\
\hline
Band & 433.546 &  &  & $\alpha$ & $\textrm{E}_{peak}$ & $\beta$ & \multirow{2}{*}{3605/318} &  \\
\cline{3-7}
(Vc) & 482.699 &  &  & $-1.610\pm0.003$ & $527\pm9$ & $-2.46\pm0.02$ &  &  \\
\hline
Band & 482.699 &  &  & $\alpha$ & $\textrm{E}_{peak}$ & $\beta$ & \multirow{2}{*}{1992/317} &  \\
\cline{3-7}
(VIa) & 508.299 &  &  & $-1.512\pm0.003$ & $564\pm8$ & $-2.55\pm0.02$ &  &  \\
\hline
Band & 515.467 &  &  & $\alpha$ & $\textrm{E}_{peak}$ & $\beta$ & \multirow{2}{*}{3677/335} & \multirow{2}{*}{Y} \\
\cline{3-7}
(VIc) & 546.188 &  &  & $-1.484\pm0.002$ & $1133\pm11$ & $-3.53\pm0.04$ &  &  \\
\hline
Band & 546.188 &  &  & $\alpha$ & $\textrm{E}_{peak}$ & $\beta$ & \multirow{2}{*}{3111/318} &  \\
\cline{3-7}
(VII) & 597.389 &  &  & $-1.648\pm0.004$ & $280\pm6$ & $-2.23\pm0.01$ &  &  \\
\hline

\hline
\hline
    \caption{The best fitting spectral functions for the lightcurve intervals and sub-intervals described in Section\,\ref{sec:temporal_analysis} and presented in Section\,\ref{sec:prompt_emission_phases}. \enquote{Y$^{*}$} is used to denote regions where the \lat \lle data spans a shorter time interval then the \gbm \cspec data due to the differing \lat and \gbm BTI regions. $\alpha$ is the low-energy photon index, $\beta$ is the high-energy photon index, and Index refers to the additional PL component index. For the mBB model, $m$ is the shape parameter described in \citealt{Hou2018} and $K$ is defined as $L_{39}/D^2_{L,10 \rm{kpc}}$, where $L$ is the luminosity in the rest frame in units of $10^{39}$\,erg s\texp{-1} and $D_{L}$ is the luminosity distance in units of 10\,kpc. All values of $\textrm{E}_{peak}$ and kT are in keV. Interval VIII (the afterglow) is not included in this table.}
\end{longtable*}


\noindent
component peaking at \abt19\,keV. The model spectrum for this fit is shown in the top left panel of Figure\,\ref{fig:ppu}.


\subsection{Pre-Main Pulse} \label{subsec:pre-main_pulse}

The bulk of the main emission episode begins at \t0+176\,s with a subdominant pulse from \t0+176\,s to \t0+210\,s (Figure \ref{fig:main}, interval III). Taken by itself, this would be one of the brightest GRBs in the \gbm sample in terms of peak flux. Although part of the main pulse, we analyzed this pulse separately because it is clearly distinct from the following bright main emission and is the last contiguous emission interval before the onset of \gbm data issues. This interval is best fit by a Band function, the parameters of this fit can be found in Table\,\ref{tab:time-res_spec_fits}.

\subsection{Primary Pulse} \label{subsec:primary_pulse}

We define the main pulse (Figure\,\ref{fig:main}, interval IV) as the region between \t0+210\,s to \t0+324\,s. The bulk of the main emission occurred during the first \gbm BTI, during which both \gbm and \lat experienced data issues\textsuperscript{\ref{ss:btis}}. Due to these data issues we divide this interval into three sub-intervals, before (IVa), during (IVb), and after (IVc) the first \gbm BTI (sub-intervals shown in Figure\,\ref{fig:main}). Although the \lat BTI falls into sub-intervals IVa and IVc, we do not attempt to correct or use any \lat \lle data within the published \lat BTIs.

Region IVa begins before the onset of \gbm data issues and is best fit by a Band function that peaks in energy (\Epk) around 3.7\,MeV. Region IVc begins after the \gbm data issues have subsided and is best fit by a Band function with an additional PL component, extending the fit out to higher energies. The Band function in this region peaks in energy (\Epk) at \abt1.4\,MeV. The additional PL of this fit has photon index of \abt-1.9, which is consistent with the canonical PL value of $\Gamma$= -2 expected from the high-energy component of the electron synchrotron spectrum for both the slow- and fast- cooling regime, for an assumed power-law electron energy distribution of $p=2$ \citep{Granot2002}. This spectral component is consistent with the emergence of the early afterglow, over which the rest of the prompt emission is superimposed, similar to behaviour observed in GRB\,190114C \citep{Ajello2020}. The model spectrum for this fit can be found in Figure\,\ref{fig:ppu}.

Region IVb is the time of the first \gbm BTI. As mentioned previously, the \tte data within the \gbm BTIs is irrecoverable. Although the \cspec data experienced PPU and deadtime effects, the data are corrected via the method described in Section\,\ref{subsec:BTIs}. Within this BTI the lightcurve consists of two distinct peaks. Our assumption of a Band function for the underlying spectral shape produces adequate fits throughout. However, we observe varying goodness-of-fit measures that coincide with the times of these two peaks. Examples of PPU-corrected counts spectra during these peaks are shown in Figure\,\ref{fig:ppu}. Due to the current limitations of our PPU-correction technique described in Section\,\ref{subsec:BTIs} we are unable to discern whether these variations are due to spectrotemporal evolution in the data or uncorrected PPU effects. Spectral fit parameters are not reported for fits within the \gbm BTI regions because the purpose of these fits is to determine the energetics of \grb. Further analysis is needed to determine the reliability of our PPU-correction technique for spectral modeling of this GRB with additional spectral models being considered.


\subsection{Intra-Pulse Period} \label{subsec:intra-pulse_period}

The intra-pulse period (region V) consists of smoother emission with three distinct pulses in sub-regions Va, Vb, and Vc as shown in Figure\,\ref{fig:main}. We fit each pulse independently, using \lle data when available. All three pulses were best fit with Band functions. Although the pulses in sub-intervals Va and Vb have similar temporal structures and MVTs (see Figure\,\ref{fig:mvt}), the pulse in sub-interval Va peaks at a much lower energy (\Epk=\abt69\,keV and \Epk=\abt520\,keV, respectively). The pulse in sub-interval Vc occurs right before the onset of the second high intensity region and has a MVT and temporal structure that clearly differs from the first two pulses. Despite these differences, this pulse peaks at an energy similar to sub-interval Vb (\Epk=\abt527\,keV). Although the intensity and variability of the emission in this period is lower than the surrounding regions, at no point does the emission truly become quiescent, showing non-thermal activity throughout the interval. 


\subsection{Secondary Pulse} \label{subsec:secondary_pulse}

Due to the secondary pulse (Figure\,\ref{fig:main}, region VI) containing the second \gbm BTI within it, we separate this region into three sub-regions (VIa, VIb, and VIc). As with the primary pulse, sub-regions VIa and VIc were not affected by \gbm data issues so standard spectral analyses could be performed.

Although the first (VIa) and last (VIc) sub-regions were both best fit by a Band function with comparable low-energy photon indices $\alpha$ of \abt-1.5 and high-energy photon indices $\beta$ values of \abt-2.5 and \abt-3.5 respectively, they differed largely in peak energy measurements. Sub-region VIa peaked in energy at \abt564\,keV while sub-regions VIc had a measured \Epk\,of \abt1.1\,MeV. Such a difference is expected as sub-regions VIc directly followed a period of reenergization.

Sub-region VIb is bound by the second \gbm BTI interval and contains only 7 time bins of \gbm \cspec data. Within this time the data contains only a single bright spike. We follow the same procedure as Section\,\ref{subsec:primary_pulse} and as described in Section\,\ref{subsec:BTIs}. The resulting spectral fits, residuals, and fit statistics adequately constrain the data in every time bin, but produce more residuals near the peak, as expected (Figure\,\ref{fig:ppu}). As discussed in Section\,\ref{subsec:primary_pulse}, we do not report any spectral fit parameters within this region in Table\,\ref{tab:time-res_spec_fits}.


\subsection{Final Pulse} \label{subsec:final_pulse}

The final pulse (Figure\,\ref{fig:main}, region VII) occurs from \t0+560\,s to \t0+597\,s. We define the end time of this region arbitrarily as it inevitably merges with the long, smooth decay phase marked as region VIII in Figure\,\ref{fig:main}. This pulse was best fit by a Band function, the parameters of which can be found in Table\,\ref{tab:time-res_spec_fits}.

\section{Afterglow} \label{sec:afterglow}

\begin{figure*}[h!tbp]
    \centering
    \includegraphics[width=\textwidth]{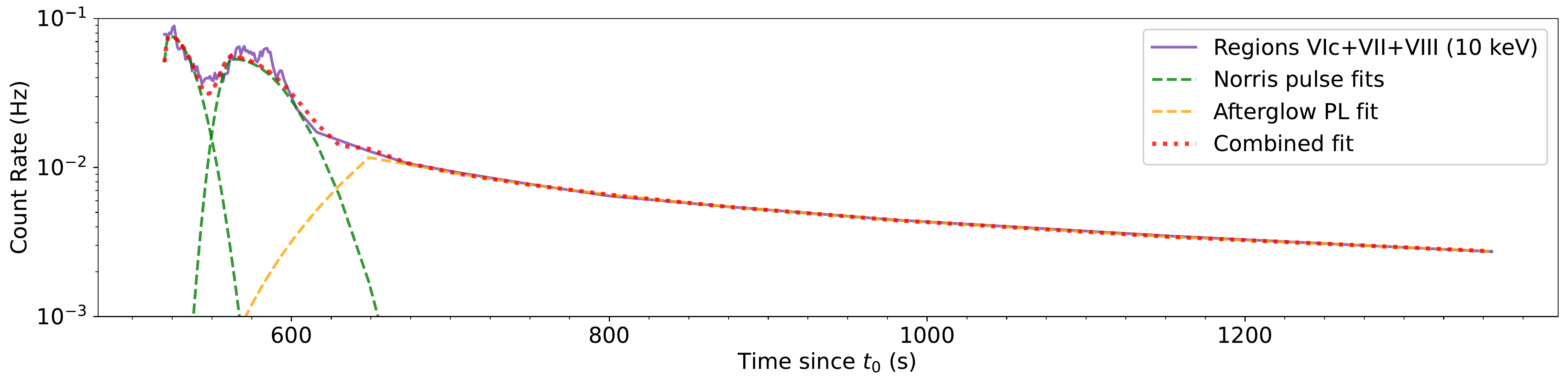}
    \includegraphics[width=\textwidth]{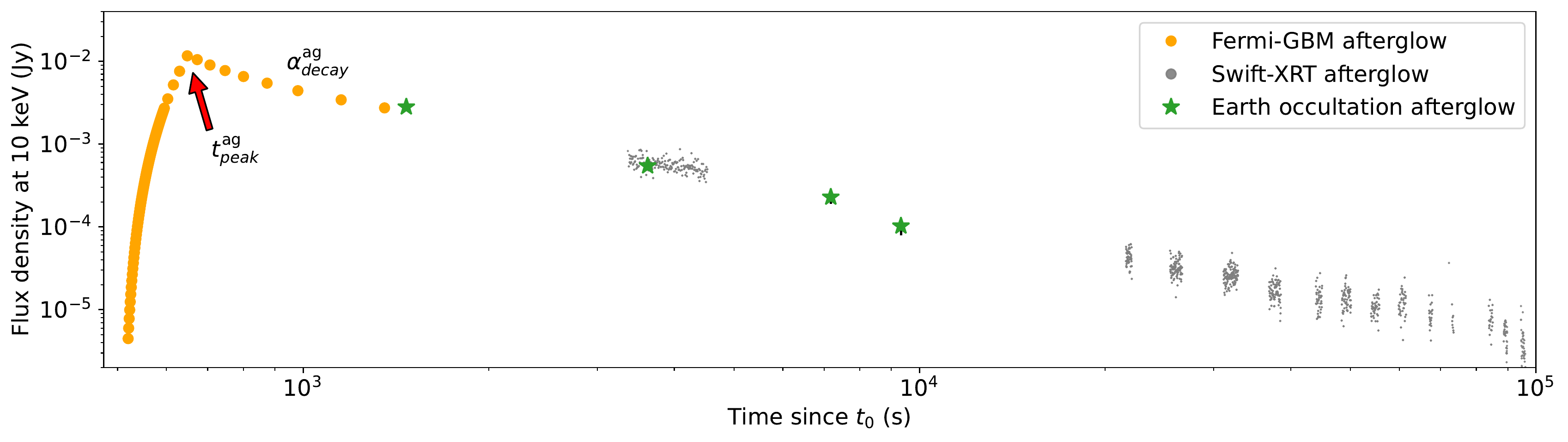}
    \includegraphics[width=\textwidth]{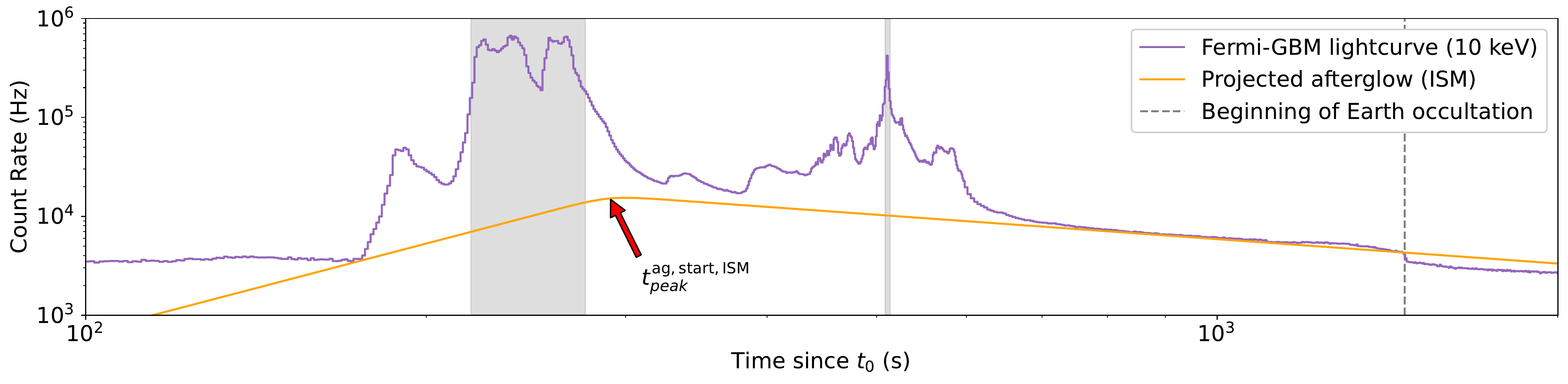}
    \caption{Top: The lightcurve of \grb in regions VIc, VII, and VIII after fitting the spectra with Band functiond and extrapolating down to 10\,keV. The green dashed lines show Norris function fits to the prompt emission phase while the yellow dashed line shows a broken power law fit to the afterglow phase. Middle: The flux density lightcurve of the afterglow (yellow dots) after removing the prompt emission flux from the Norris function pulses. Flux measurements with the Earth occultation technique are also shown by green stars and the gray points mark the \swift-XRT observations. Bottom: The full \gbm lightcurve at 10\,keV fitted with the projected ISM-afterglow such that the afterglow emission does not exceed the prompt emission. The region to the right of the vertical dashed line ($t_{0}+1464$\,s) marks the time when \grb was occulted by the Earth.}
    \label{fig:afterglow}
\end{figure*}

In the case of some GRBs the afterglow is so bright it can contribute detectable flux in the \gbm band pass \citep{Giblin1999, Connaughton2002}. As seen in GRB\,190114C and discussed in Section\,\ref{subsec:primary_pulse}, this afterglow flux can also overlap with the highly variable prompt emission. For \grb, the last discernible pulse (region VII) peaks around \t0+575\,s and is followed by a long, smooth decay period (region VIII). This decay period having no appreciable variability (see Figure\,\ref{fig:mvt}) is consistent with an afterglow origin. We define the end of this period to be \t0+1460\,s, the time when \grb is occulted by Earth for \fermi. 

We temporally bin the lightcurve from \t0+510\,s (small bump in region VIc) to the end of region VIII, requiring a signal to noise ratio of 80 to ensure an adequate amount of signal. We then fit the spectrum in each resulting interval with a Band function. Next, we extrapolated the spectrum down to 10\,keV (Figure\,\ref{fig:afterglow}, purple) for comparison with \swift-XRT \citep{Williams2023}. 

In order to constrain the peak of the afterglow we fit the 10\,keV extrapolated lightcurve with two \citet{Norris2005} pulse models (Figure\,\ref{fig:afterglow}, green lines) for the prompt emission and a broken power-law (Figure\,\ref{fig:afterglow}, yellow line) for the afterglow region. The gamma-ray afterglow lightcurve will rise with a power law index of 3 (ISM) or 1/2 (wind), then decay as a power law (different indices for rising are possible depending on the ordering of the characteristic frequencies). We can only isolate the late decay part of the afterglow as the peak is hidden under prompt emission. The broken power-law (Figure\,\ref{fig:afterglow}, yellow dots) peaks at $t^{\rm ag, \rm ISM}_{peak}\gtrsim t_{\rm ref}+(140\pm2)$\,s and $t^{\rm ag, \rm wind}_{peak}\gtrsim t_{\rm ref}+(120\pm6)$\,s with a temporal decay slope of $\alpha^{\rm ag}_{decay}=-0.82\pm0.03$ ($t_{ref}=510$\,s). This slope is relatively shallow compared to normal afterglow decay. We tentatively identify $t^{\rm ag}_{peak}$ as the peak of onset of the solely-afterglow emission and assume the true afterglow peak is overwhelmed by the prompt emission. 

Additionally, the slope ($\alpha^{\rm ag}_{decay}$) of the following afterglow region resembles a plateau phase with mean index $\overline{\alpha^{\rm ag}_{decay}}=-0.6$ and standard deviation $\sigma_{\alpha^{\rm ag}_{decay}}=0.4$ \citep{Grupe2013}. The later emission ($t>t_{0}+1400$\,s) \swift-XRT slope of $\alpha_{s}=-1.514 \pm0.003$\footnote[6]{\url{https://www.swift.ac.uk/xrt_live_cat/01126853/}} agrees with the expected value of the afterglow ($\overline{\alpha_{s}}=-1.5$, $\sigma_{\alpha_{s}} =0.6$; \citealt{Grupe2013}).

We obtain different constraints for the peak of the afterglow by extrapolating the smooth emission of phase VIII back in time (Figure\,\ref{fig:afterglow}, bottom). We fixed the afterglow lightcurve temporal decay index to our measured $\alpha^{\rm ag}_{decay}$ value, set our reference time to be the start of the bulk emission ($t_{\rm ref} = t_{0}+175$\,s), and fixed our temporal rise index, $\alpha^{\rm ag}_{rise}$, to either 3 for an ISM external medium or 1/2 for a wind-type external media. Requiring that the extrapolated afterglow flux not exceed the prompt emission flux, we place a limit of $t^{\rm ag, \rm start, \rm ISM}_{peak}\gtrsim t_{\rm ref}+105 $\,s. A similar value for the wind-type external medium ($t^{\rm ag, \rm start, \rm wind}_{peak}$) is not reported because the fit was unconstrained. We note that all afterglow peak times mentioned above depend on the \citet{Norris2005} pulse and broken power-law functional fits to our 10\,keV extrapolated lightcurve and therefore have an additional small yet uncharacterized uncertainty associated with them.

\gbm can also measure source fluxes using an Earth occultation technique, modeling the change in count rate when a source of interest goes behind the Earth or emerges from behind the Earth \citep{WilsonHodge2012}. Using the known \swift-XRT sky localization for \grb, Earth occultation times were first estimated for the time of 50\% transmission at 100\,keV. A 240\,s data window surrounding each occultation step was fitted using a model comprised of a quadratic background and source terms; in this case for \grb and Cygnus X-1. Independent fits were performed for each energy channel and each detector viewing \grb within 60 degrees from the detector normal. The source terms consist of an energy dependent model of the atmospheric transmission convolved with the detector response and an assumed spectral model. Two fixed spectral models were tested, a single power-law with a photon index of -2, and a Band function with $\alpha=-0.85$, $\beta=-2.0$, and $E_{\textrm{peak}}=21.0$\,keV based on the afterglow analysis. Because each energy channel is fitted independently, these models only apply across a single \gbm \ctime or combined set of \gbm \cspec channels, so results between the two models were consistent. The four Earth occultation steps with excesses above background starting 1470\,s after the trigger are shown in Figure\,\ref{fig:afterglow} as green stars. The afterglow fluxes estimated from the Earth occultation steps are consistent with the temporal decay observed by \swift-XRT.

The Earth occultation analysis for \grb differs slightly from that described in \citet{WilsonHodge2012} because the usual focus of the \gbm occultation technique is on the study of longer-term variations (days to years) rather than individual steps. Data filtering, normally used to reject highly variable backgrounds, was omitted for these fits because the time-interval of interest was getting rejected due to the brightness of \grb. Only Cygnus X-1, the brightest potentially interfering source, was fitted within the time windows rather than the full catalog of flaring sources in \citet{WilsonHodge2012} to avoid rejecting the data of interest.

\section{Lorentz Factor} \label{sec:lorentz_factor}

With the requirement that the emission site be optically thin, we can derive lower limits on the bulk Lorentz factor ($\Gamma^{\rm ot}_{min}$; \citealt{Lithwick2001}). For this method we usually use the MVT in conjunction with the spectral shape. Based on our COMP function spectral fit in region Ia of the triggering pulse ($\alpha=-1.69\pm0.01$, $E_{\rm peak}=3.98\pm0.37$\,MeV, and energy flux=$(1.98\pm0.03)\times 10^{-6}$ erg cm$^{-2}$ s$^{-1}$ in the 10-1000 keV range), we find a bulk Lorentz factor of $\Gamma^{\rm trig, \rm ot}_{min}\gtrsim 188$. 

Deriving the bulk Lorentz factor in the main prompt emission in the same way is more difficult because we can only derive the MVT for time intervals without PPU. However, the spectrum within the \gbm BTIs yields much stronger constraints on $\Gamma^{\rm ot}_{min}$ even with conservative assumptions for the MVT. Based on Figure\,\ref{fig:mvt} and the discussions in Section\,\ref{subsec:mvt}, we assume a conservative MVT estimate of $\Delta t_{\rm var}=0.1 $\,s. Again requiring that the prompt emission be optically thin and utilizing the PPU-corrected spectrum in the brightest time interval we set a limit of $\Gamma^{\rm prompt, \rm ot}_{min}\gtrsim 1040$. If we instead assume $\Delta t_{\rm var}=0.05 $\,s, which is still reasonable given the MVT values surrounding region IV, we achieve a Lorentz factor limit of $\Gamma^{\rm prompt, \rm ot}_{min}\gtrsim 1470$. 

We can also set single zone Lorentz factor limit through pair-production for the highest energy photons ($\Gamma^{\rm pp}_{min}$). \lat observed a 99.3 GeV photon at \t0+240\,s \citep{gcn32658}. Assuming the \gbm emission emanates from the same volume and the requirement that photons can escape without producing $e^{\pm}$ pairs yields a Lorentz factor constraint of $\Gamma^{\rm prompt, \rm pp}_{min}\gtrsim 1560$. We note that the calculation of $\Gamma^{\rm pp}_{min}$ by \citealt{Lithwick2001} is done under the assumption of a single zone. If we instead consider a more realistic situation, taking into account the angular, temporal, and spatial dependence of the radiation field, our values of $\Gamma^{\rm pp}_{min}$ could be lower by a factor of \abt2 (i.e., $\Gamma^{\rm prompt, \rm pp}_{min}\gtrsim 780$) \citep{Hascoet2012, Gill2018, Vianello2018, Arimoto2020}. Although both MVT-derived optically thin Lorentz factor lower limits fall below the single zone pair-production Lorentz factor lower limit, the two methods independently produce consistent results.

The peak of the afterglow emission denotes the beginning of the external shock deceleration time. By identifying this peak for both ISM and wind-type external media as $t^{\rm ag, \rm ISM}_{peak}\gtrsim t_{0}+(140\pm1.5)$\,s and $t^{\rm ag, \rm wind}_{peak}\gtrsim t_{0}+(120\pm6.5)$\,s respectively, we can derive upper limits for the Lorentz factor of the afterglow via,

\begin{equation} \label{eq:Lorentz_factor_general}
    \Gamma^{\rm ag}_{min} = \pp{\p{\frac{(17-4s)}{16\pi(4-s)}}\p{\frac{E_{k}}{n_0 m_{\rm p}c^{5-s}}}}^{\frac{1}{8-2s}}\p{\frac{t_{\rm peak}}{(1+z)}}^{-\frac{3-s}{8-2s}}
\end{equation}

\noindent
where $s=0$ assumes an ISM external density profile and $s=2$ assumes a wind-type external medium \citep{Nappo2014, Ghirlanda2018}. For both cases we use the VLT reported redshift of z = 0.151 \citep{gcn32648} and assume the kinetic energy, $E_k$ is approximately equal to the isotropic-equivalent gamma-ray energy ($E_{k}\approx E_{iso}$).

Assuming an ISM external density profile ($s=0$) we have:

\begin{equation} \label{eq:Lorentz_factor_ISM}
    \Gamma^{\rm ag, \rm ISM}_{min} \gtrsim 260 \left(\frac{E_{k,55}}{n/1~ {\rm cm}^{-3}}\right)^{1/8} \left(\frac{t_{\rm peak}}{120 ~{\rm s} \times 1.151}\right)^{-3/8}
\end{equation}
Using the $t^{\rm ag, \rm start, \rm ISM}_{peak} = 105 $\,s value, the Lorentz factor increases slightly to $\Gamma^{\rm ag, \rm start, \rm ISM}_{min}\gtrsim 270$ with all scaling parameters being the same.

For the wind-type external medium ($s=2$) the density parameter, $n= 3\times 10^{35}~A_\star~ {\rm cm}^{-1}$ \citep{grb221009a_fermi_lat_collaboration_2023}, gives us:

\begin{equation} \label{eq:Lorentz_factor_wind}
    \Gamma^{\rm ag, \rm wind}_{min}\gtrsim 282 \left(\frac{E_{k,55}}{A_{\star,-1}}\right)^{1/4} \left(\frac{t_{\rm peak}}{ 140 ~{\rm s} \times 1.151}\right)^{-1/4}
\end{equation}

\section{Energetics} \label{sec:energetics}

\begin{figure*}[h!tbp]
    \centering
    \includegraphics[width=\textwidth]{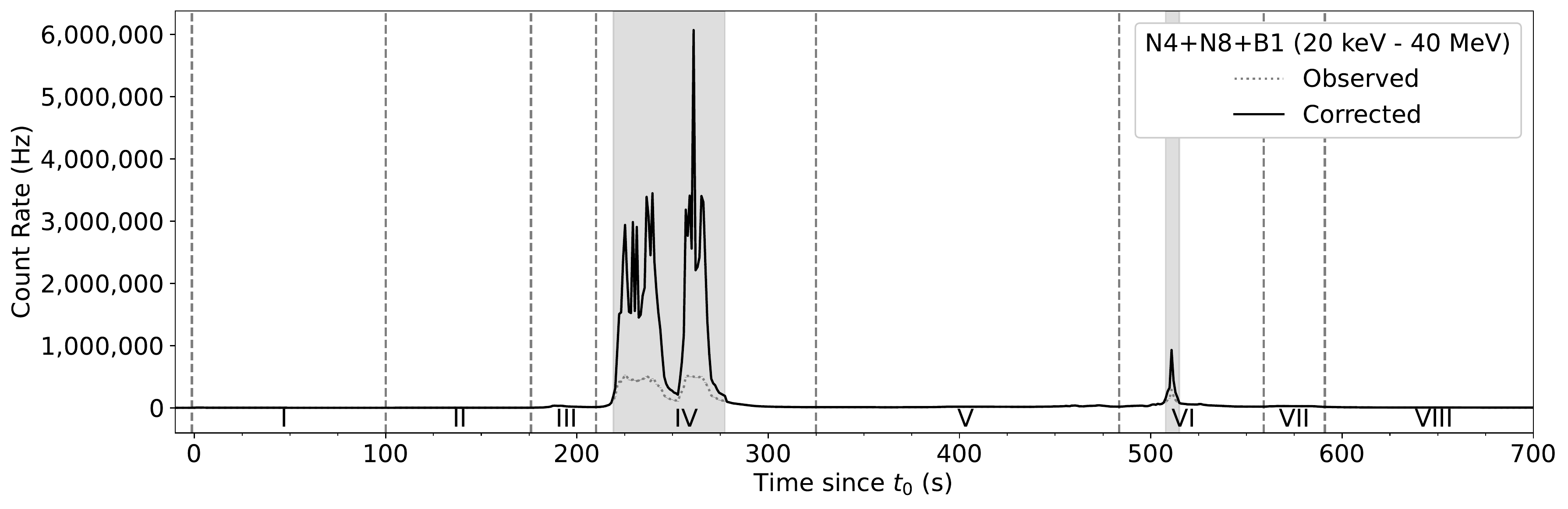}
    \begin{minipage}[b]{0.49\linewidth}
        \includegraphics[width=\textwidth]{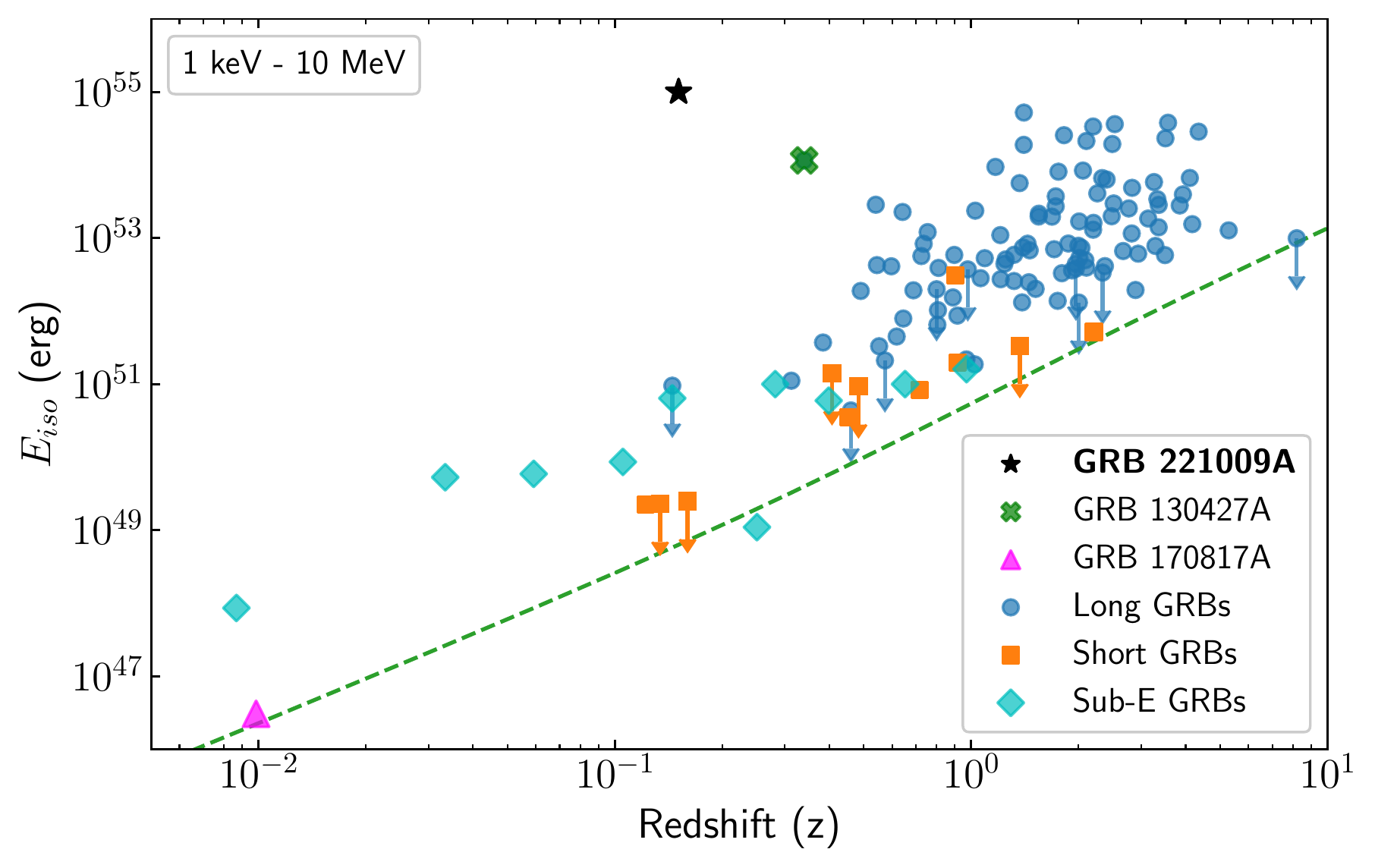}
    \end{minipage}
	\begin{minipage}[b]{0.49\linewidth}
        \includegraphics[width=\textwidth]{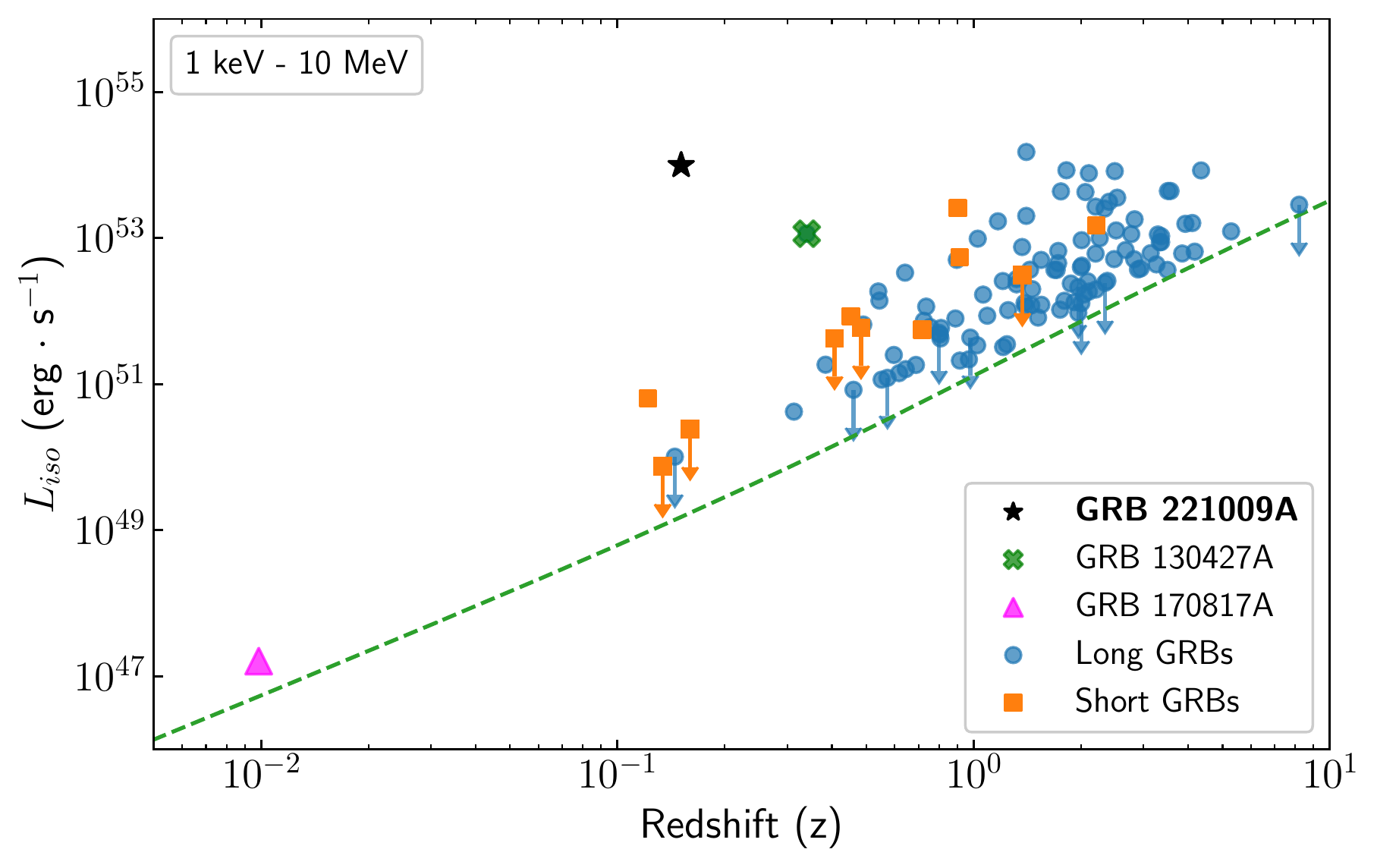}
    \end{minipage}
    \caption{Top: The lightcurve of \grb from 8\,keV to 40\,MeV before and after PPU-correcting in the \gbm BTI region. Bottom: Short and long GRB $E_{\rm iso}$ (left) and $L_{\rm iso}$ (right) measures of \gbm-detected GRBs with known redshift through 2017 \citep{Abbott2017a}, with updated spectral measures from \citet{Poolakkil2021}. Key GRBs are highlighted. The downward arrows indicate GRBs best fit by a PL spectral model, which must turnover somewhere, so they are shown as upper limits. The $E_{\rm iso}$ plot is supplemented with low or intermediate luminosity GRBs with associated supernova from \citet{Cano2017}. We do not utilize their luminosity measures as they are averaged (not peak).}
    \label{fig:energetics}
\end{figure*}

In order to calculate burst energetics the PPU-corrected data within the two \gbm BTI regions must be used. As was done with the $\textrm{T}_{90}$ analysis, all non-BTI regions shown in Figure\,\ref{fig:main} were also fit with a  Band function. For intervals where a Band function is not the preferred spectral form, we find it still produces an adequate fit and therefore introduces negligible errors in our results.

We derive the total isotropic equivalent energy ($E_{\gamma,\textrm{iso}}$) in the 1-10,000\,keV range from the fluences in the individual time intervals from \t0-2.7\,s to \t0+1449.5\,s and perform k-corrections in all intervals. We find the total Fluence = $(9.47\pm0.07)\times10^{-2}$\,erg\,cm$^{-2}$ and $E_{\gamma,\textrm{iso}}=(1.01\pm0.007)\times10^{55}\,\erg$. Assuming an opening angle of 2 degrees \citep{Negro2023}, the beaming corrected energy is $E_{\gamma}=6.1\times10^{51} \left(\theta_j/2\right)^2$\,erg. We obtain the isotropic-equivalent luminosity, by integrating the spectrum in the \t0+230.8\,s to \t0+231.8\,s interval. The energy flux here is $F=(8.48 \pm 0.06)\times10^{-2}$\,erg s$^{-1}$\,cm$^{-2}$. After k-correction, we find the 1\,s peak luminosity to be $L_{\gamma,\textrm{iso}}=(9.91 \pm 0.06)\times 10^{53}$\,erg s$^{-1}$. These values are consistent with those independently produced by \citet{Frederiks2023}, \citet{Ripa2023}, and \citet{An2023} which demonstrate the validity of the \gbm PPU-correction technique for this GRB.

\section{Interpretation} \label{sec:interpretation}

With the discovery of GRB\,211211A, a long GRB ($\textrm{T}_{\rm 90}=34.2\pm0.6$\,s; \citealt{Veres2023}) being more consistent with a compact binary merger origin rather than a collapsar origin \citep{Gompertz2023, Rastinejad2022, Troja2022, Yang2022}, the class of short GRBs with extended emission is now being reconsidered. This class of GRBs was first introduced after GRB\,060614 was detected by \swift-BAT \citep{DellaValle2006, Gal-Yam2006, Gehrels2006}. This means \grb, being a long GRB due to its $\textrm{T}_{90}$ duration of $289\pm1$\,s (Section\,\ref{subsec:duration}), is no longer sufficient evidence for associating its progenitor to a massive star.


\subsection{Central Engine} \label{subsec:central_engine}

Comparison of the triggering pulse (Figure\,\ref{fig:main}, region I) against the orbital-averaged background (Figure\,\ref{fig:background_osv}) suggests a near full return to background around \t0+100\,s, followed by weak signal at \t0+121\,s (Figure\,\ref{fig:main}, region II) before the onset of the main pulse. The triggering pulse of \grb is unique due to its isolation in time, outlier spectral parameter values, no detectable emission prior to its onset, and \lle photons leading the \gbm data, possibly pointing to a distinct physical origin.

The best fit model for the triggering pulse is a mBB which is suggestive of photospheric thermal emission arriving at the observer from different locations on the equal arrival time surface \citep{Peer2008, Deng2014}. The COMP model, which fits the triggering pulse equally well, can be generated via the mBB model when $m=0$. Although the value of $m$ in Table\,\ref{tab:time-res_spec_fits} is not zero, it is still reasonably close to reproducing the COMP spectral shape.

For a spherically emitting shell, the MVT relates the radius of the shell to the Lorentz factor via,
\begin{equation} \label{eq:min_radius}
    \textrm{R} \sim \Delta t_{\textrm{min}} \p{\frac{\Gamma^{2} c}{\p{1+z}}}
\end{equation}
\noindent
where R is the radius of the spherically emitting shell, $\Gamma$ is the Lorentz factor, $c$ is the speed of light, and $z$ is the measured redshift. Along these lines, the MVT and the bulk Lorentz factor can be used to place limits the on emitting shell radius, assuming a singular emitting region. In interval Ia, the average value of MVT is \abt0.1\,s. Using the afterglow-derived bulk Lorentz factor for the ISM ($\Gamma^{\rm ag, \rm ISM}_{min} \gtrsim 260$) and the wind-type external medium ($\Gamma^{\rm ag, \rm wind}_{min} \gtrsim 282$) as a proxy for the bulk Lorentz factor in this region gives us an estimate on the radius of the emitting star. We find $R^{\rm ISM}_{\star} \gtrsim 1.7 \times 10^{14}$\,cm and $R^{\rm wind}_{\star} \gtrsim 2.0 \times 10^{14}$\,cm for the ISM and wind-type media respectively. If we instead use the Lorentz factor derived for this region ($\Gamma^{\rm trig, \rm ot}_{min}\gtrsim 188$), we achieve an estimated radius of $R^{\rm trig}_{\star} \gtrsim 9.0 \times 10^{13}$\,cm. All of which are roughly consistent with the radius of the outer wind from the Wolf-Rayet progenitor star ($R_{\star} > 1 \times 10^{13}$\,cm) \citep{Crowther2007}.

With the triggering pulse having thermally dominant spectral properties, a clear and distinct start time, a derived emitting shell radius consistent with that of a Wolf-Rayet progenitor star, along with the discovery of the associated supernova SN 2022xiw \citep{gcn32800, gcn32850, Fulton2023, Srinivasaragavan2023}; \grb is likely the result of a massive core-collapse supernovae progenitor. Although the beam-corrected $E_{\gamma}$ value of $6.1\times10^{51} \left(\theta_j/2\right)^2$\,erg from Section\,\ref{sec:energetics} is within the maximum energy release limit for a magnetar progenitor ($\sim3\times10^{52}$\,erg; \citealt{Usov1992}), this progenitor source is unlikely.


\subsection{Shock Breakout} \label{subsec:shock_breakout}

Shock breakout occurs when the radiation transport velocity is faster than the shock velocity and the radiation is no longer trapped within the shock \citep{Fryer2023}. This is not strictly at the stellar photosphere, but is often near it for a supernova shock and will be further out for a relativistic shock \citep{Fryer2020}. Just as with shock breakout and shock interaction for core-collapse supernovae, if the early emission is thermal, it can be used to probe characteristics of the progenitor and its immediate surroundings as well as the structure of the outflow. In the context of gamma-ray bursts, thermal emission can probe the progenitor-star photosphere (which is set by the mass loss and stellar radius), inhomogeneities of the mass-loss and structure of the jet, and its cocoon. For massive Wolf-Rayet stars, the likely progenitors of gamma-ray bursts, the high wind mass-loss rate often places the photosphere in the stellar wind and, for relativistic shocks, the shock breakout radius will be in the stellar wind. For this paper, we focus on more fundamental aspects behind a thermal component to determine whether it is a reasonable explanation of the observed emission, deferring a detailed comparison of the data to models for a later paper. For a more detailed discussion on the physics of shock breakout see \citealt{Fryer2020}.

If we assume the observed triggering emission is produced by the Lorentz-boosted thermal emission of shock breakout, the limits and shape of the emission can be used to constrain the properties of the shock as it emerges from the star. In the strong shock limit for a highly-relativistic gas, the pressure of the shock ($P_{\rm shock}$) is,

\begin{equation} \label{eq:P_shock}
    P_{\rm shock} \approx \Gamma^2 \rho_{\rm CSM} c^2
\end{equation}

\noindent
where $\Gamma$ is the Lorentz factor, $\rho_{\rm CSM}$ is the density in the region of shock breakout (in the wind of the massive star), and $c$ is the speed of light. Assuming the pressure is radiation-dominated, the corresponding temperature ($T_{\rm shock}$) of the emitting gas in the gas comoving frame is,

\begin{equation} \label{eq:T_shock}
    T_{\rm shock} \approx 1.9 (\Gamma/100)^{0.5} (\rho_{\rm CSM}/10^{-10}\,g\,cm^{-3})^{0.25} \; \textrm{keV}
\end{equation}

\noindent
and the corresponding peak energy of the emitted photons ($\nu_{\rm peak}$) in the observer frame is,

\begin{equation} \label{eq:nu_shock}
    \nu_{\rm peak} \approx 1 (\Gamma/100)^{1.5} (\rho_{\rm CSM}/10^{-10}\,g\,cm^{-3})^{0.25} \; \textrm{MeV}
\end{equation}

In the thermal shock breakout paradigm, the broad range and relatively flat spectra for the prompt emission requires a distribution of Lorentz factors. The observed peak emission around 15\,MeV places strong constraints on the upper limit of the Lorentz factor, corresponding to peak Lorentz factors lying between 300--1000, with corresponding densities of $70$--$0.05 \times 10^{-10} \, {\rm g \, cm^{-3}}$.


\subsection{Prompt Emission} \label{subsec:prompt_emission}

Unlike the triggering pulse, the spectra for the bulk of the prompt emission (regions III through VII) are all best fit with a Band function. Unlike the mBB and COMP spectral models, the Band function can not be generated through the superposition of Planck-like spectra, pointing to the bulk of the prompt emission having a non-thermal origin. Using Equation \ref{eq:min_radius}, the lower limit of the shock breakout derived triggering pulse Lorentz factor ($\Gamma=300$), and a delay time of \abt220\,s between regions I and III, we get an internal dissipation radius of \abt$6\times10^{17}$\,cm. This value is larger than is typically discussed but could be consistent with the ICMART model of $10^{16}$\,cm, which is consistent with the non-detection of neutrinos \citep{gcn32665, gcn32741} and a Poynting flux dominted jet \citep{Zhang2010}. The transition from thermal to non-thermal emission is not unique amongst GRBs. But this transition occurring in two isolated emission episodes separated by long quiet period has only ever been seen in one other burst, GRB\,160625B \citep{Zhang2018}. However, what is unique about \grb is we can, for the first time, directly say that from onset to afterglow the central engine never appears to shut off throughout the duration of the GRB.


\subsection{Afterglow} \label{subsec:afterglow}

As discussed in Section\,\ref{sec:afterglow}, we estimate the afterglow onset time to be $\sim t_{0}+280 $\,s. This time is at the end of the first \gbm BTI region, but it is possible that the afterglow began within the first \gbm BTI region itself. This aligns both with the -2 power-law spectral component fitted in region IVc and with the detection of the 99\,GeV \lat photon, further reinforcing that the afterglow begins during the prompt emission phase and the two are seen in a superposition with one another. Furthermore, We were also able to isolate a time at which the emission is comprised of solely-afterglow which provides us with a distinct time at which the central engine ceased emitting.

Comparing the highest and lowest Lorentz factors derived for the prompt emission ($\Gamma^{\rm prompt, \rm pp}\sim1560$ and $\Gamma^{\rm prompt, \rm pp}\sim780$ after correcting for a factor of 2) to those of the afterglow ($\Gamma^{\rm ag, \rm wind}\sim282$ and $\Gamma^{\rm ag, \rm ISM}\sim260$) yields a deceleration of $\Delta\Gamma^{\rm ISM}\sim1300$ and $\Delta\Gamma_{\rm wind}\sim1278$ (or $\Delta\Gamma_{\rm ISM}\sim520$ and $\Delta\Gamma_{\rm wind}\sim498$) over \abt380\,s. This rapid deceleration is suggestive of a relativistic reverse-shock (i.e., a \enquote{thick shell}) and is consistent with the deceleration time not exceeding the prompt GRB duration.


\subsection{Uniqueness} \label{subsec:uniqueness}

We therefore consider if a triggering pulse like this would be identified in other GRBs. Some bright long GRBs have weak trigger intervals followed by quiescence before the main emission episode, but most do not. The 1.024\,s peak-flux interval of this pulse would trigger \gbm out to z\,$\approx$\,1.3. Approximately half of \gbm GRBs with measured redshifts are within this range. Thus, a similar pulse should have been recovered in other bursts. The particularly high $\Gamma$ for this burst would produce a higher luminosity shock breakout, which may explain the lack of identification in other collapsars.

Figure\,\ref{fig:energetics} places the isotropic energy and luminosity into context with a broad sample of \gbm and \lat bursts with known redshift through 2017 \citep{Abbott2017a}, with updated spectral measures from \citet{Poolakkil2021}. \grb stands alone. It is the highest $E_{iso}$ in the \gbm sample, and with no higher claim existing in the literature, it is the brightest $E_{iso}$ ever recorded for a GRB. It is four orders of magnitude higher than GRBs seen at comparable redshifts and is an order of magnitude greater than the \enquote{Nearby Ordinary Monster} GRB\,130427A\, making \grb a nearerby extraordinary monster. In $L_{\rm iso}$, only GRB\,160625B is marginally higher; however, effects due to PPU and calibration issues caused by the geometry of incident photons with respect to \fermi make this our most limited measure.

\section{Summary} \label{sec:summary}

Our dedicated search suggests that there is no emission from the central engine prior to the \gbm trigger time. Weak emission from the triggering pulse (region I) can be localized to \grb for up to 100\,s before fully returning to background. The first \abt8\,s of the triggering pulse has a peak energy (\Epk) of 4\,MeV with the highest energy photons arriving first. The entire triggering pulse has spectrotemporal properties best characterized by the Lorentz-boosted thermal emission of shock breakout. Using the peak energy of the assumed shock breakout emission we are able to place the Lorentz factor ($\Gamma$) of this pulse between 300 and 1000 with corresponding shock densities between 70 and 0.05$\times 10^{-10}\, {\rm g \, cm^{-3}}$. These properties along with our derived emitting shell radius and the discovery of the associated supernova SN 2022xiw  point to a core-collapse Wolf-Rayet star being the most likely central engine progenitor.

We find another weakly emitting pulse (region II) which we were able to localize to \grb just before the bulk of the main emission began. The central engine is then continuously active, showing non-thermal activity throughout the bulk of the prompt emission. With two periods of severe re-brightening, at no point does the prompt emission become quiescent. After correcting for PPU effects in the \gbm data we were able to characterize the total energetics of this burst. With a total isotropic-equivalent energy of $\textrm{E}_{\gamma,\textrm{iso}}=1\times10^{55}$\,erg and an isotropic-equivalent luminosity of $\textrm{L}_{\gamma,\textrm{iso}}=9.9\times10^{53}$\,erg s$^{-1}$, \grb is the most intrinsically energetic and second most intrinsically luminous in the \gbm sample. Assuming a MVT of 0.05\,s we place constraints on the Lorentz factor during the brightest interval via pair-creation using the \lat reported 99.3\,GeV photon and by assuming an optically thin environment. These values are $\Gamma^{\rm prompt, \rm pp}_{min}\gtrsim 1560$ and $\Gamma^{\rm prompt, \rm ot}_{min}\gtrsim 1470$ respectively (without the factor of 2 correction).

By comparing our measurements with those observed by \swift-XRT we are able to characterize the onset of the early-afterglow period. We place lower limits on the afterglow peak time assuming both ISM and wind-type external media of $t^{\rm ag, \rm ISM}_{peak}\gtrsim t_{0}+(140\pm1.5)$\,s and $t^{\rm ag, \rm wind}_{peak}\gtrsim t_{0}+(120\pm6.5)$\,s respectively. We additionally find an early plateau region with a slope $\alpha^{ag}_{decay}=-0.82\pm0.03$ which gradually steepens to the observed \swift-XRT slope of $\alpha_{s}=-1.5$ at \t0$>$1400\,s. We are also able to project this afterglow back into the prompt emission, where the two are in a superposition with one another, and estimate the start time of the afterglow ($t^{\rm ag, \rm start, \rm ISM}_{peak}\gtrsim t_{0}+280$\,s). Again assuming both an ISM and wind-type external media we calculate the Lorentz factor of the external shock as $\Gamma^{\rm ag, \rm ISM}_{min} \gtrsim 260$ and $\Gamma^{\rm ag, \rm wind}_{min} \gtrsim 282$ respectively. The change in Lorentz factor from prompt emission to afterglow is suggestive of a relativistic reverse-shock.

The observations presented here provide an unrivaled probe into the continuously active central engine. Without the abundance of photons provided by this GRB, the lower flux intervals (e.g., the triggering pulse, between the \gbm BTIs, and the afterglow) would have been lost amongst the background, creating a \enquote{tip of the iceberg} effect with only the high flux intervals being visible. While we could infer the continuity of the central engine in other GRBs, we have never had the abundant photons to observe the central engine as well as was done for \grb. While the spectroscopy of the triggering pulse is strongly indicative of a thermal, photospheric origin, as is expected to occur early in GRBs, its significantly lower intensity compared to the rest of the emission makes its detection in other bursts particularly challenging. Furthermore, the abundance of photons in \grb allows us to track the evolution of the bulk Lorentz factor through to the afterglow phase, providing a stronger indication that the reverse shock is encountered as the emission enters the afterglow phase.


\vspace{5mm}

\noindent We dedicate this paper to the memory of William \enquote{Bill} Paciesas who passed away in June 2022. Bill was a co-investigator of BATSE and, for a time, the Principal Investigator of \gbm, but he was arguably more well known by the high-energy astrophysics community for his punny humor and beer connoisseurship. With \grb being the Brightest Event Ever Recorded (BEER) and Bill not here to share it with us; this BEER's for you.

\vspace{5mm}

\section{Acknowledgments} \label{sec:acknowledgments}

S.L. acknowledges useful discussions with Tyson Littenberg and thanks him for the assistance with writing the Bayesian Markov Chain fitting technique. The UAH co-authors gratefully acknowledge NASA funding from co-operative agreement 80MSFC22M0004. The USRA co-authors gratefully acknowledge NASA funding through contract 80MSFC17M0022. D.K., C.A.W.H., and C.M.H. gratefully acknowledge NASA funding through the Fermi GBM project. Support for the German contribution to GBM was provided by the Bundesministerium f¨ur Bildung und Forschung (BMBF) via the Deutsches Zentrum f¨ur Luft und Raumfahrt (DLR) under contract number 50 QV 0301. R.H. acknowledges funding from the European Union’s Horizon 2020 research and innovation programme under the Marie Skłodowska-Curie grant agreement No 945298-ParisRegionFP.

The \lat Collaboration acknowledges generous ongoing support from a number of agencies and institutes that have supported both the development and the operation of the LAT as well as scientific data analysis. These include the National Aeronautics and Space Administration and the Department of Energy in the United States, the Commissariat \`a l'Energie Atomique and the Centre National de la Recherche Scientifique / Institut National de Physique Nucl\'eaire et de Physique des Particules in France, the Agenzia Spaziale Italiana and the Istituto Nazionale di Fisica Nucleare in Italy, the Ministry of Education, Culture, Sports, Science and Technology (MEXT), High Energy Accelerator Research Organization (KEK) and Japan Aerospace Exploration Agency (JAXA) in Japan, and the K.~A.~Wallenberg Foundation, the Swedish Research Council and the Swedish National Space Board in Sweden. ECF is supported by NASA under award number 80GSFC21M0002.

Additional support for science analysis during the operations phase is gratefully acknowledged from the Istituto Nazionale di Astrofisica in Italy and the Centre National d'\'Etudes Spatiales in France. This work performed in part under DOE Contract DE-AC02-76SF00515.

This work made use of data supplied by the UK Swift Science Data Centre at the University of Leicester.

\appendix
\section{Fermi Time-Resolved Interval Selection} \label{app:fermi_intervals}

\fermi provides spectral coverage over seven orders of magnitude in energy for short duration transients, beginning around 8\,keV in \gbm and reaching into the hundreds of GeV with the \lat. Time-resolved analyses using \fermi data can map the evolution of distinct spectral components in time. For most GRBs, the \gbm \tte data are binned to match the intervals of interest determined by burst structure or the counts in a given data type. Due to the various data issues caused by \grb, this burst requires special care when determining the appropriate temporal intervals to use. 

The selections must first account for the burst duration exceeding the pointing stability timescale of \fermi. Visibility by \gbm to the source position begins 2111\,s before trigger time (\t0) and lasts until \t0+1550\,s when the region is occulted by the Earth. The visibility for the \lat is limited by the \lat field of view. The source leaves the \lat primary field of view at \t0+380\,s and exits the wider \lle data field of view at \t0+600\,s. 

Data selections must also account for the distinct intervals of \gbm and \lat with and without pulse pile-up, and the capability to correct for pulse pile-up effects. In \gbm these intervals are from approximately \t0+219\,s to \t0+277\,s and from approximately \t0+508\,s to \t0+514\,s. Within these time intervals \tte data were lost due to the limited bandwidth of the \gbm electronics and are irrecoverable. \ctime and \cspec data within these intervals exist but are severely effected by pulse pile-up and are unreliable in their current, uncorrected state. For regions surrounding the \gbm intervals with pulse pile-up, the temporal boundaries to switch between \tte and binned \cspec (or \ctime) data are based on the temporal binning and phase of the binned data. This ensures the transition occurs on the edges of the binned data. Using the bin edges that fully cover the intervals with pulse pile-up gives selections times from \t0+218.501\,s to \t0+277.894\,s and from \t0+507.275\,s to \t0+514.443\,s.

Throughout the duration of \grb both the standard \lat data product and the \lat \lle data experienced data issues within certain time regions. Although we leave the details of these effected time regions to the \lat paper (CITE LAT PAPER) we would like to mention how to properly handle using both data products simultaneously. \lle data are generated at 0.1\,s and 1.0\,s temporal resolution with their phase set to the bin boundary at \gbm trigger time. For \gbm intervals without pulse pile-up, the \gbm \tte data can be binned to match the edges of the \lle data. This is not possible during \gbm intervals with pulse pile-up because \tte data are irrecoverable during these times. Therefore, pulse pile-up corrected \cspec or \ctime data must be used. For regions when the \gbm \cspec or \ctime data are out of phase with the \lle data, the \lle data must be re-binned to be in-phase with the \gbm binned data. A similar re-binning procedure must also be done over the edges of these time intervals with either \tte or \lle data to handle the transitions both in and out of the \gbm intervals with pulse pile-up.

\begin{table}[h!tbp]
\begin{tabular}{cccc}
\hline
\hline
$t_{\rm start}$ & $t_{\rm stop}$ & \tte & \cspec (or \ctime) \\
\hline
-10.000 & 210.000  & Y & Y       \\
210.000 & 218.501  & Y & Y       \\
218.501 & 277.894  &   & Y$^{*}$ \\
277.894 & 290.000  & Y & Y       \\
290.000 & 380.000  & Y & Y       \\
380.000 & 507.275  & Y & Y       \\
507.275 & 514.443  &   & Y$^{*}$ \\
514.443 & 600.000  & Y & Y       \\
600.000 & 1500.000 & Y & Y       \\
\hline
\hline
\end{tabular}
\caption{The \gbm data products valid within a given time interval. All time range values are relative to the \gbm trigger. The data types with an asterisk represent \gbm data types with pulse pile-up and should only be used for scientific interpretation after correcting for pulse pile-up effects.}
\end{table}

\bibliographystyle{aasjournal}


\end{document}